\documentclass[showkeys, reprint, superscriptaddress, amsmath, amssymb, aps, prb, floatfix]{revtex4-2}
\usepackage[export]{adjustbox}
\usepackage{graphicx}% Include figure files%
\usepackage[font=small,justification=justified,format=plain]{caption}
\usepackage[labelformat=empty, position=top, justification=justified,format=plain]{subcaption}
\usepackage{xcolor}

\usepackage{dcolumn}% Align table columns on decimal point
\usepackage{bm}% bold math
\usepackage{siunitx}
\usepackage{hyperref}% add hypertext capabilities
\usepackage[mathlines]{lineno}% Enable numbering of text and display math
\usepackage{titlesec}
\usepackage{multirow}

\titlespacing\section{0pt}{16pt plus 1pt minus 1pt}{5pt plus 1pt minus 1pt}
\titlespacing\subsection{0pt}{16pt plus 0.2pt minus 0.2pt}{5pt plus 0.2pt minus 0.2pt}
\titlespacing\subsubsection{0pt}{12pt plus 0.2pt minus 0.2pt}{0.2pt plus 0.2pt minus 0.2pt}

\makeatletter %remove space between cit number [1,2]
\def\NAT@def@citea{\def\@citea{\NAT@separator}}
\def\fnum@figure{\figurename\thefigure}
\renewcommand\eqref[1]{\textup{Eq.\kern-0.4ex~(\ref{#1})}} %format distance between Eq.~and number
\makeatother

\usepackage{lipsum, babel}
\usepackage{blindtext}
\usepackage{hyperref}
\usepackage{xcolor}
\hypersetup{ %cool links and refs
    colorlinks = True,
    allcolors = blue
}
\newcommand{\rom}[1]{\uppercase\expandafter{\romannumeral #1\relax}}

\DeclareSIUnit\angstrom{\text{Å}}

\usepackage{microtype}

\begin{document}

\preprint{APS/123-QED}

\title{Electron Mobilities in SrTiO$_3$ and KTaO$_3$: Role of Phonon Anharmonicity, Mass Renormalization and Disorder}

\author{Luigi Ranalli\thanks}
\thanks{Email: luigi.ranalli@univie.ac.at}
    \affiliation{Faculty of Physics and Center for Computational Materials Science, University of Vienna, Kolingasse 14-16, 1090 Vienna, Austria}
    \affiliation{Vienna Doctoral School in Physics, University of Vienna, Boltzmanngasse 5, 1090 Vienna, Austria}

\author{Carla Verdi}%
    \affiliation{School of Mathematics and Phisycs, University of Queensland, 4072 Brisbane, Queensland, Australia}
\author{Marios Zacharias}%
    \affiliation{Univ Rennes, INSA Rennes, CNRS, Institut FOTON - UMR 6082, F-35000 Rennes, France}
\author{Jacky Even}%
    \affiliation{Univ Rennes, INSA Rennes, CNRS, Institut FOTON - UMR 6082, F-35000 Rennes, France}
\author{Feliciano Giustino}%
    \affiliation{Oden Institute for Computational Engineering and Sciences, The University of Texas at Austin, Austin, Texas 78712, USA}
    \affiliation{Department of Physics, The University of Texas at Austin, Austin, Texas 78712, USA }
\author{Cesare Franchini}%
    \affiliation{Faculty of Physics and Center for Computational Materials Science, University of Vienna, Kolingasse 14-16, 1090 Vienna, Austria}
    \affiliation{Department of Physics and Astronomy ”Augusto Righi” Alma Mater Studiorum - Università di Bologna, 40127 Bologna Italy}

\begin{abstract}
Accurately predicting carrier mobility in strongly anharmonic solids necessitates a precise characterization of lattice dynamics as a function of temperature. We achieve consistency with experimental electron mobility data for bulk KTaO$_3$ and SrTiO$_3$ above 150~K by refining the Boltzmann transport equations. This refinement includes incorporating temperature-dependent anharmonic phonon eigenfrequencies and eigenmodes into the electron-phonon interaction tensor, while maintaining the derivatives of the Kohn-Sham potential as computed in density functional perturbation theory. Using efficient machine-learned force fields and the stochastic self-consistent harmonic approximation, we accurately compute the dynamical matrices. At room temperature, the calculated mobility for SrTiO$_3$ exceeds experimental values by an order of magnitude, whereas the overestimation for KTaO$_3$ is way less pronounced. This discrepancy is explained through the more significant electron mass renormalization near the conduction band bottom due to anharmonic electron-phonon coupling and the presence of local disorder in SrTiO$_3$.
\end{abstract}


\maketitle

\section{INTRODUCTION}
\label{Introduction}
Transition-metal oxide perovskites are a widely studied class of materials due to their broad spectrum of intriguing physical and chemical properties such as high-temperature superconductivity~\cite{Bednorz1988,Reyren2007}, magnetoresistance~\cite{Jin1994}, ferroelectricity~\cite{Cohen1992}, multiferroicity~\cite{Wang2009,Narayan2018} and piezoelectricity~\cite{Li2018}, arising from  
their structural and chemical versatility and from the intermingled coupling of orbital, spin, lattice, and charge degrees of freedom~\cite{Imada1998,He2012,Ergonenc2018,Varrassi2021,Liu2024}.~These properties allow for novel applications in non-volatile memory devices, spintronics, nanoelectronics and energy technologies, like solid oxide fuel cells and metal-air batteries~\cite{Dogan2015, Sawa2008, Sun2021, Kim2014, Suntivich2011}.

Understanding charge transport in these materials is of utmost importance for virtually all applications, as it can shed light on the underlying mechanisms behind experimental measurements and aid in optimizing carrier mobility.~From a computational perspective, solving the Boltzmann transport equation (BTE) for charge carriers with \textit{ab initio} predicted parameters is a powerful strategy that is gaining increasing popularity due to its accuracy and to the recent availability of efficient implementations~\cite{ponce2018, Protik2020a,Park2020,Zhou2021,Gonze2020}.
% and the increase in computational resources available, 
Despite the abundant literature on standard semiconductors~\cite{Restrepo2009, Zhou2016, Zhao2020b, Protik2020b, Ponce2019a, ponce2019b, ponce2020, ponce2020b, ponce2021,  Macheda2018, Ma2018a, Ma2018b, Liu2017, Liu2018, Li2015, Lee2018, Lee2020, Kang2019, Jhalani2020, Cao2018, Brunin2020b}, the calculation of carrier mobility in transition metal perovskites is comparatively limited~\cite{Zhou2018, Zhou2019,Ma2022,Wang2023,Shen2023}.

As discussed in previous works~\cite{ponce2018,ponce2021}, moderate deviations of the BTE prediction from the experimental mobility point at an incomplete treatment of the electronic correlation, dielectric screening, and/or carrier effective masses. Semi-empirical re-scaling techniques or more accurate many-body calculations can be shown to mitigate these discrepancies.~A persisting deviation from experimental measurements, instead, could indicate either the inadequacy of the harmonic electron-phonon coupling approximation, or the breakdown of the quasi-particle (QP) approximation on which the BTE relies, in favour of more complex transport mechanisms.~For example, strong electron-phonon coupling may lead to electronic satellites near the conduction band bottom and significant broadening of the electronic bands, which both break the QP picture.~Such features have been found in angle-resolved photoemission spectroscopy (ARPES) spectra of SrTiO$_3$ surfaces~\cite{Wang2016}, and the presence of polaronic satellites has been proposed as an explanation of both optical conductivity sub-bands~\cite{Mechelen2008,Klimin2020} in bulk SrTiO$_3$ and mobility reduction in the 150-300~K temperature range~\cite{Zhou2019} using 
the Kubo formula.~Despite the similar conduction band orbital character, ARPES spectra of KTaO$_3$ surfaces do not show any polaronic peak~\cite{Santander-Syro2012}, even though significant QP broadening in bulk KTaO$_3$ cannot be excluded at room temperature.

In the electron-phonon scattering regime the experimental mobility in bulk KTaO$_3$ is almost one order of magnitude greater than in SrTiO$_3$~\cite{Cain2013,Wemple1965}.~To elucidate the origin of this rather surprising observation, we perform BTE calculations to compute the electron mobilities in both materials in the range between 150 and 300~K.~In order to account for the strong temperature dependence of the phonon dispersions, we renormalize the electron-phonon matrix elements by introducing accurate anharmonic dynamical matrices calculated by using the stochastic self-consistent harmonic approximation (SSCHA) method~\cite{Hooton1955,Errea2013,Errea2014,Bianco2017,Monacelli2018,Monacelli2021}.~Leveraging accurate machine-learned force fields (MLFFs) we are able to access large supercells in SSCHA, employing the scheme recently proposed for these two compounds~\cite{Verdi2023, Ranalli2023}.~Such approach is strictly necessary for computing the mobility in SrTiO$_3$, since its phonon dispersions present imaginary modes at the harmonic level. The inclusion of temperature-dependent SSCHA dynamical matrices is shown to better recover the temperature dependence of the mobility in both materials. 

Interestingly, the computed mobilities are comparable on the whole temperature range, and largely overestimate the experimental ones, especially in the case of SrTiO$_3$.~After examining dielectric screening corrections, spin-orbit coupling effects, the role of anharmonic phonons and the effect of electronic correlation on the effective masses, we propose the greater electron-phonon renormalization of the effective masses at the conduction band bottom in SrTiO$_3$ as compared to KTaO$_3$ as the reason for the massive mobility overestimation. This effect is not considered in standard BTE calculations, where the electronic bands are generally computed using density functional theory (DFT) 
%using local or semilocal exchange-correlation functionals
with the nuclei clamped to their equilibrium positions. The discrepancy is further mitigated by the additional mass renormalization brought by the presence of local disorder in SrTiO$_3$~\cite{Zacharias2023}, which we predict to be absent in KTaO$_3$.
Finally, we propose the QP broadening and the potential insufficiency of the harmonic electron-phonon coupling approximation as possible explanations for the residual overestimation of experimental mobility, which remains higher than what is typically found in other semiconductors.~\cite{ponce2021}. 

\section{\label{sec:computational_methods}COMPUTATIONAL METHODS}
The strategy adopted for the renormalization of the electron-phonon matrix elements with the inclusion of temperature-dependent anharmonic phonon effects is described in Section~\ref{sec:Renormalization of the Electron-Phonon Matrix Elements}; the training of MLFFs~\cite{Jinnouchi2019}  used for capturing the correct phonon dispersions through the SSCHA is explained in Section~\ref{sec:Training of MLFFs for SSCHA}; and the inclusion of local disorder in the carrier mobility calculations is described in Section~\ref{sec:Mass Enhancement and Inclusion of Local Disorder}.

\subsection{Renormalization of the Electron-Phonon Matrix Elements and Mobility Calculations}
\label{sec:Renormalization of the Electron-Phonon Matrix Elements} 
The calculations of the electron mobilities via the BTE are performed using the EPW code~\cite{ponce2016,Giustino2007,Verdi2015,Lee2023}, based on the formalism summarized in Supplemental Material (SM) Section~\rom{1}~\cite{SM}.~For this purpose, we need to calculate the electron-phonon matrix elements on coarse \textbf{k}- and \textbf{q}- Brillouin-zone (BZ) grids:
\begin{equation}
\label{eq:g}
    g_{mn\nu}(\mathbf{k},\mathbf{q}) = \langle u_{m\mathbf{k}+\mathbf{q}} | \Delta_{\mathbf{q}\nu} v^{\rm KS} | u_{n\mathbf{k}} \rangle_{uc}
\end{equation}
where $u_{n\mathbf{k}}$ is the Bloch-periodic component of the Kohn-Sham (KS) electron wavefunction and the integral is performed over the unit cell.~The $u_{n\textbf{k}}$ are computed within DFT using Quantum Espresso (QE)~\cite{Giannozzi2009,Giannozzi2017}.~The $\Delta_{\mathbf{q}\nu} v^{\rm KS}$ operator is the variation of the KS potential with respect to the collective displacement associated to a phonon mode $\nu$ with wavevector $\textbf{q}$, and it is in turn expressed as:
\begin{equation}
\label{eq:dvks}
    \Delta_{\mathbf{q}\nu} v^{\rm KS} = l_{\mathbf{q}\nu} \sum_{\kappa\alpha} (M_0/M_\kappa)^{1/2} e_{\kappa\alpha,\nu}(\mathbf{q}) \partial_{\kappa\alpha,\mathbf{q}} v^{\rm KS}
\end{equation}
where $l_{\mathbf{q}\nu}$=[$\hbar / (2M_0 \omega_{\mathbf{q}\nu})$]$^{1/2}$ is the zero-point displacement amplitude, $M_0$ a reference mass [that cancels out in Eq.~(\ref{eq:dvks})], $M_\kappa$ the mass of atom \textit{$\kappa$}, $\omega_{\mathbf{q}\nu}$ the phonon eigenfrequency for branch $\nu$, and $e_{\kappa\alpha, \nu}$ the eigenvector in the Cartesian direction $\alpha$.~In the presence of imaginary $\omega_{\mathbf{q}\nu}$, as is the case of SrTiO$_3$, $\Delta_{\mathbf{q}\nu} v^{\rm KS}$ is not well defined.~As a consequence, the electron-phonon matrix elements and the corresponding computed mobilities show spurious behaviors.~Furthermore, in the standard harmonic approximation the temperature dependence of the phonon frequencies is neglected, thus leading to an incomplete estimate of $\Delta_{\mathbf{q}\nu} v^{\rm KS}$ in strongly anharmonic materials, as is the case for both KTaO$_3$ and SrTiO$_3$.~In order to overcome these problems, we calculate $\partial_{\kappa\alpha,\mathbf{q}} v^{\rm KS}$ within standard density functional perturbation theory (DFPT) using QE on a 4$\times$4$\times$4 reciprocal \textbf{q}-grid, but we substitute $e_{\kappa\alpha,\nu}(\mathbf{q})$ and $\omega_{\mathbf{q}\nu}$ as obtained by diagonalization of the anharmonic dynamical matrices calculated on the same 4$\times$4$\times$4 reciprocal \textbf{q}-grid (see Section~\ref{sec:Training of MLFFs for SSCHA}).~We thus proceed with the calculation of $g_{mn\nu}(\mathbf{k},\mathbf{q})$ for each temperature of interest by introducing the associated temperature-dependent anharmonic $e_{\kappa\alpha,\nu}(\mathbf{q})$ and $\omega_{\mathbf{q}\nu}$.~The comparison of the electron-phonon matrix elements obtained in this way and from harmonic phonons is shown in SM Fig.~S1~\cite{SM} for both compounds (for more details, see SM Section~\rom{2}~\cite{SM}).~Lastly, the standard Wannier rotations are performed in order to interpolate the electron Hamiltonian, phonon dynamical matrix and electron-phonon matrix elements from the coarse \textbf{k}- and \textbf{q}-grids to the fine reciprocal-space grids required for the convergence of the mobility calculations~\cite{Giustino2017} (equations shown in detail in SM, Section~\rom{1}~\cite{SM}).~For this purpose, EPW employs the Wannier90 package~\cite{Pizzi2020} for obtaining maximally localized Wannier functions in real space~\cite{wannier1937, Marzari1997}. 

The electron mobility in KTaO$_3$ and SrTiO$_3$ is computed with and without spin-orbit coupling (SOC) using the relaxation time approximation (RTA) and iterative BTE (IBTE) approximations to the standard BTE.~For both materials, a 100$\times$100$\times$100 \textbf{k}- and \textbf{q}-grid is employed for the fine-grid interpolation of the electron-phonon matrix elements to compute the mobilities.~Doping concentrations of $n = 0.8 \times 10^{17} \, \text{cm}^{-3}$ for SrTiO$_3$ and $n = 2.4 \times 10^{18} \, \text{cm}^{-3}$ for KTaO$_3$ are employed, in order to match the doping concentration of the experimental samples~\cite{Cain2013,Wemple1965}. Nevertheless, in the 150-300~K temperature range, the experimental mobility is independent from doping concentration.~The Hall mobility is also evaluated in a $10^{-10} \, \text{T}$ magnetic field.

The mobility calculations for the disordered supercell structure (shown in SM Fig.~S2~\cite{SM}) are executed by further substituting in the BTE equations the new electron bands extracted from the disordered supercell (shown in SM Fig.~S3~\cite{SM}).~More details can be found in Sec.~\ref{sec:Mass Enhancement and Inclusion of Local Disorder} and SM Section~\rom{3}~\cite{SM}.

The DFT and DFPT calculations are performed on the cubic phase of KTaO$_3$ and SrTiO$_3$ at the experimental lattice constants of, respectively,~$\SI{3.9842}{\AA}$ and~$\SI{3.900}{\AA}$, thus neglecting lattice expansion with temperature.~We use the Perdew-Burke-Ernzerhof (PBE)~\cite{Perdew1997} functional and fully relativistic norm-conserving pseudopotentials from PseudoDojo~\cite{Hamann2013}, adopting a 8$\times$8$\times$8 Monkhorst-Pack k-point mesh and a plane-wave cut-off of 140~Ry.
%and an electronic convergence threshold of $10^{-16}$ Ry.
The PBE functional is chosen for both KTaO$_3$ and SrTiO$_3$ since the DFPT formalism is not yet implemented at the more accurate meta generalized gradient approximation (meta-GGA) and random-phase approximation (RPA) levels.~Meta-GGA and RPA calculations are instead used in order to obtain accurate anharmonic dynamical matrices (see Section~\ref{sec:Training of MLFFs for SSCHA}).

\subsection{Training of MLFFs for SSCHA anharmonic phonon calculations}
\label{sec:Training of MLFFs for SSCHA}
We obtain the anharmonic dynamical matrices as a function of temperature via SSCHA, aided by the MLFF method implemented in the Vienna Ab initio Simulation Package (VASP)~\cite{Kresse1993,Kresse1996}.~Using this method, accurate kernel-based machine-learning models of the potential energy surface of a material, as well as its derivatives (forces, stress tensor) can be constructed.

The MLFF for KTaO$_3$ is built through on-the-fly {\it ab initio} molecular dynamics (MD).~The MD runs are performed at fixed volume and temperature (NVT ensemble) at 100, 200, 300 and 400~K with a time step of 2~fs and using the final structure at each temperature as the starting configuration for the higher ones.~A Langevin thermostat is employed~\cite{Hoover1982}, with a friction coefficient of 10~ps$^{-1}$ for each atomic species.~The cut-off radius for both the two- and three-body atomic density distribution descriptors is set to $\SI{13}{\AA}$, and the Gaussian broadening to 0.3~\AA.~421 {\it ab initio} training configurations were gathered over a total of 50,000 MD steps.~The {\it ab initio} calculations are performed at the meta-GGA level using the Strongly Constrained and Appropriately Normed (SCAN) functional~\cite{Sun2015} and Projector Augmented Wave (PAW) potentials~\cite{Kresse1999}, with a plane-wave cutoff of 800~eV.~We adopt a hard pseudo-potential for the Oxygen atoms, and include semicore s-states for K and p-states for Ta, thus keeping the same number of electrons treated as valence as in the QE pseudo-potentials.~A cubic 4$\times$4$\times$4 supercell and a 2$\times$2$\times$2 Monkhorst-Pack k-point mesh are employed.
%We set an electronic convergence threshold of $10^{-6}$ eV.~

The training data for SrTiO$_3$ are available from Ref.~\cite{Verdi2023}.~Generating the dataset at the RPA level was shown to be necessary in order to correctly reproduce the experimental phonon dispersions as a function of temperature via SSCHA.~In the present work, the MLFF is re-fitted using the same training dataset and methodology and employing a larger 13~\AA\ cut-off radius for the atomic density distributions, as done for KTaO$_3$~\cite{Ranalli2023}.~This ensures that the optical phonon dispersions near the $\Gamma$-point are correctly reproduced after the inclusion of the non-analytical term correction~\cite{Pick1970} (see SM Fig.~S4(a)~\cite{SM}).

The trade-off between enhanced cut-off radii and error on the ground-state energies is investigated in SM Section~\rom{4}~\cite{SM}, and the validation tests for SrTiO$_3$ and KTaO$_3$ are reported, respectively, in SM Fig.~S4(b) and Fig.~S5~\cite{SM}.~Additional details on the MLFF+SSCHA calculations are reported in SM Section~\rom{5}~\cite{SM}.

\subsection{Mass Enhancement and Inclusion of Local Disorder}
\label{sec:Mass Enhancement and Inclusion of Local Disorder} 
Local disorder or polymorphism significantly influences the electronic structure of many perovskite materials~\cite{Zhao2020a, Zacharias2023}, which in turn is expected to impact their mobilities.~To account for the effect of local disorder in our calculations, we employ the methodology described in Refs.~\cite{Zacharias2023prb, Zacharias2023} implemented in the ZG special displacement package of the EPW code~\cite{Lee2023}.~We consider 2$\times$2$\times$2 supercells of SrTiO$_3$ and KTaO$_3$ and displace the atoms along the soft modes starting from the respective high-symmetry cubic structures.~The amplitude of these displacements was set equal to the zero-point displacement by setting the associated $\omega_{\mathbf{q}\nu}$ equal to its absolute value.~We then perform a geometry optimization by allowing the nuclei to relax, keeping the lattice constants fixed.~In the 2$\times$2$\times$2 supercell, we find locally disordered networks yielding an energy lowering of 9~meV per unit cell for SrTiO$_3$, while no energetically favourable disordered ground-state structure is found for KTaO$_3$.~This result highlights the presence of an evident degree of local disorder in SrTiO$_3$.~To demonstrate how local disorder affects the electronic structure of SrTiO$_3$ in the first BZ, we compute the electron spectral function using a band structure unfolding technique for plane waves~\cite{Popescu2012, Zacharias2020} and the ZG package~\cite{Lee2023}.

To calculate the band gap renormalization and effective mass enhancements at finite temperatures, we employ the special displacement method (SDM)~\cite{Zacharias2020} in 6$\times$6$\times$6 supercells relying on the anharmonic dynamical matrices computed in SSCHA (see Section~\ref{sec:Training of MLFFs for SSCHA}).~We calculate the mobility effective mass within the SDM as the harmonic mean~\cite{ponce2021}:
\begin{equation}
\label{eq:harmonic_mean}
    m = \frac{3m_{x}m_{y}m_{z}}{m_{x}m_{y}+m_{x}m_{z}+m_{y}m_{z}}
\end{equation}
where $m_{x,y,z}$ are obtained by finite differences of the energy at the $\Gamma$-point considering the average of the three lowest eigenvalues at the conduction band minimum (CBM). The effective masses reported in Table~\ref{tab:mass_enhancements} are calculated via Eq.~(\ref{eq:harmonic_mean}).

\section{RESULTS AND DISCUSSION}
\label{sec:results}
\subsection{SrTiO$_3$}
\label{sec:STO}
\begin{figure}[t]
        \includegraphics[width=0.5\textwidth, trim={0.4cm 0cm 0cm 0.5cm},clip]{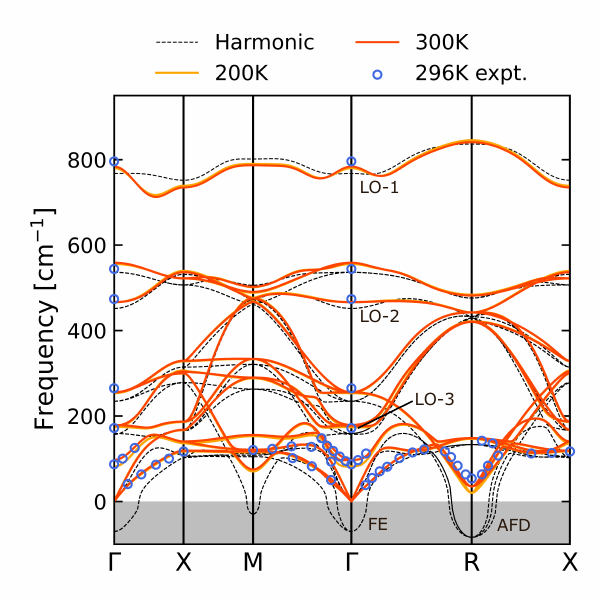}
        \vspace{-0.5cm}
        \caption{\label{fig:STO_phonons}Temperature-dependent anharmonic phonon dispersion (solid lines) and the harmonic phonon dispersion (dashed lines) in SrTiO$_3$.~Experimental data $\SI{296}{K}$ are also shown (circles)~\cite{Servoin1980,Stirling1972}.}
\end{figure}

\begin{figure}[t]
        \includegraphics[width=0.5\textwidth, trim={0.4cm 0.1cm 0.1cm 0.3cm},clip]{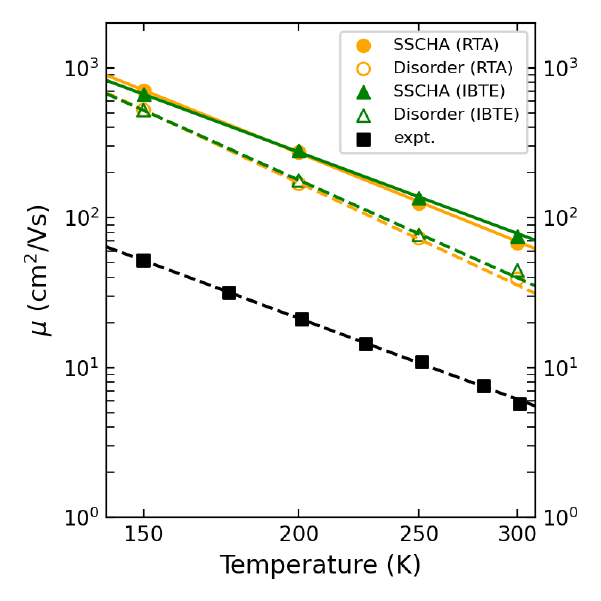}
    \vspace{-0.5cm}
    \caption{\label{fig:STO_mobility}Electron mobility calculations in SrTiO$_3$ using SSCHA temperature-dependent dynamical matrices substitution. The BTE is solved via RTA (orange filled circles) and IBTE (green filled triangles).~Local disorder is included by further substitution of the high-symmetry unit cell electronic eigenvalues with those of the disordered supercell at RTA (orange empty triangles) and IBTE (green empty triangles) level.~Both the calculated and the experimental mobilities~\cite{Cain2013} are fitted using Eq.~(\ref{eq:power_law}).~All the computed mobilities are rescaled by Eq.~(\ref{eq:rescale}).} 
\end{figure}

\begin{figure*}[htbp]
    \includegraphics[width=\linewidth, trim={0cm 0.0cm 0cm 0cm},clip]{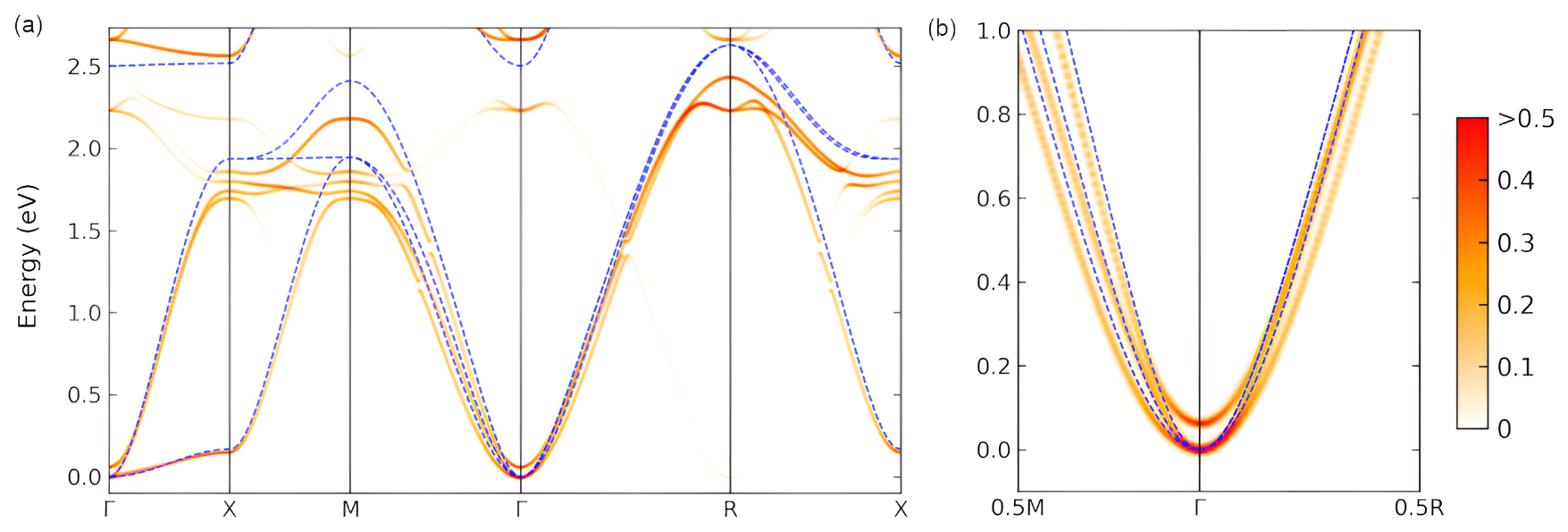}
    \caption{(a) Electron spectral function (colour map) of SrTiO$_3$ resulting from unfolding the bands of the 2$\times$2$\times$2 disordered supercell onto the high-symmetry path of the cubic unit cell.~The band structure for the symmetric unit cell is shown as dashed blue line.~ Both CBMs have been shifted to zero.~(b) Zoom on the 0.5M-$\Gamma$-0.5R path, where only energies close to the CBM are shown.}
    \label{fig:F3}
\end{figure*}

The anharmonic phonon dispersion near room temperature as calculated through MLFF+SSCHA shown in Fig.~\ref{fig:STO_phonons} is in excellent agreement with the experimental dispersion measured at 296~K~\cite{Servoin1980,Stirling1972}.~At the harmonic level, SrTiO$_3$ exhibits imaginary frequencies for the ferroelectric (FE) and antiferrodistortive (AFD) lowest energy modes, which lead to spurious divergences of the electron-phonon matrix elements (see SM Fig.~S1(a)~\cite{SM}).~In order to obtain reasonable mobility calculations within the harmonic approximation, we introduce a phonon cut-off frequency $\omega_\textup{cut}$, and thus set $g_{mn\nu}(\mathbf{k},\mathbf{q})$ to zero for $\nu$ and $\mathbf{q}$ such that $\omega_{\mathbf{q}\nu} \leq \omega_\textup{cut}$.~A cut-off of 10~meV is thus chosen, since the mobility converges to stable values in its vicinity.~With this setup, we observe an overestimation of the electron mobility in the lower end of the 150-300~K temperature range, where these modes have a non negligible interaction with the charge carriers (see SM Fig.~S7~\cite{SM}).~Apart from the rather small interaction with the FE and AFD modes, the less temperature-sensitive LO-1, LO-2 and LO-3 modes are the main scattering channels, with LO-1 interacting over the whole BZ and the LO-3 mode being strongly interacting only around the $\Gamma$-point, as shown in SM Figs.~S1(a)-(b)~\cite{SM}.

At low temperatures, charge carriers scatter mainly by lattice impurities, while at higher temperatures electron-phonon scattering dominates the dynamics and the Hall mobility is expected to follow a power law:
\begin{equation}
\label{eq:power_law}
    \mu_{\textup{H}} = A \times T^{-\gamma} .
\end{equation}
A fit to the experimental data reproduced in Fig.~\ref{fig:STO_mobility} yields $\gamma=3.08$.~Adopting the harmonic approximation results in a clear deviation %from of the calculated data 
from this power law, as shown in SM Fig.~S7~\cite{SM}.~By employing the anharmonic SSCHA matrices instead, the expected trend is recovered, as shown in Fig.~\ref{fig:STO_mobility}.~The IBTE calculations yield $\gamma=3.07$, in very close agreement with the experimental value, while the RTA yields a slightly overestimated $\gamma=3.34$.~SrTiO$_3$ thus broadly follows the universal scaling law $\mu \propto T^{-3.3}$ for halide perovskites obtained by considering the LO-1 mode as the only scattering channel~\cite{ponce2019c}.~These results are in close agreement with previous calculations~\cite{Zhou2018}.~Note that including the temperature dependence of the anharmonic dynamical matrices is important.~Neglecting their temperature evolution by fixing them to the ones calculated at 150~K leads to an underestimation of the computed mobility (see SM Fig.~S8~\cite{SM}), since the hardening of the FE and AFD modes with increasing temperature is not taken into account and the scattering is overestimated as predicted by Eq.~(\ref{eq:dvks}).

The overestimation of the calculated electron mobility for SrTiO$_3$ compared to experimental values is substantially more pronounced than that observed in previous BTE calculations for standard semiconductors~\cite{ponce2018}.~To address this issue, we explore various effects that may be responsible for such overestimation.~The inclusion of SOC leads to a 28~meV splitting of the conduction bands near the CBM and slight changes in the electronic effective masses (see SM Fig.~S9 and Table S\rom{1}~\cite{SM}).~Hence, SOC has little to no effect on the mobility according to a simplified transport model~\cite{Himmetoglu2016} and as confirmed by our calculations (see SM Fig.~S10~\cite{SM}).~Employing a more accurate account of electronic correlation effects trough the HSE06 hybrid functional brings little to no change in the effective masses~\cite{janotti2011}.~Our computed electronic dielectric constant of 6.17, though, overestimates the experimental value of 5.62~\cite{Merker1955}, a common issue in DFT calculations.~By rescaling the mobility as~\cite{ponce2021}:
\begin{equation} \label{eq:rescale}
     \mu^\textup{screen} = \mu \times (\epsilon^\textup{DFT}/\epsilon^\textup{exp})^2,
\end{equation}
a decrease of 17\% is achieved, still not enough to correct the overestimation of more than 800\% with respect to the experimental value at 300~K.

An important phenomenon so far neglected is the presence of local disorder, that changes the equilibrium position of the atoms in each unit cell without affecting the global lattice symmetry.~By relaxing a 2$\times$2$\times$2 supercell of SrTiO$_3$ with broken symmetry as detailed in Section~\ref{sec:Mass Enhancement and Inclusion of Local Disorder}, a new ground state is found with an energy lowering of 9~meV per unit cell (see SM Fig.~S2~\cite{SM}), in agreement with the recent work of Ref.~\cite{Zacharias2023}.~The associated unfolded band structure in Fig.~\ref{fig:F3} shows a decrease in the curvature of the conduction band bottom, which translates to an average electronic effective mass of $m^*_{\text{dis}}=0.58\,m_{e}$, a factor 1.22 greater than $m_0=0.49\,m_{e}$ obtained for the symmetric unit cell.~The resulting mobility decrease estimated through~\cite{ponce2021}:
%m0=0.58\,m_{e}$
\begin{equation} \label{eq:mass_rescale}
     \mu^* = \mu \times (m^* / m_0)^{-1}
\end{equation}
is thus of 22\%.~The electronic band gap is also strongly affected, with an increase of 0.31~eV with respect to the 1.92~eV of the symmetric unit cell.~In Table~\ref{tab:mass_enhancements} we summarize the overall mobility overestimation with respect to the experimental value at 300~K for the aforementioned and subsequent levels of approximation, along with band-gap corrections.~A splitting of 62~meV near the CBM is also observed, twice the one caused by the inclusion of SOC. 

In order to compare the approximate rescaling of the mobility via  Eq.~(\ref{eq:mass_rescale}) with a full mobility calculation on the disordered structure, we substitute the new electronic eigenvalues obtained for the 
disordered structure in the BTE (SM Eq.~(S5)~\cite{SM}), together with the previously computed SSCHA matrices. We obtain a mobility decrease of 28\% and 43\% at 150~K and 300~K, respectively, as compared to the previous estimate on the symmetric unit cell, as shown in Fig.~\ref{fig:STO_mobility}.~The trend still follows Eq.~(\ref{eq:power_law}) with~$\gamma=3.71$.~The decrease in the computed mobility is higher than the one obtained through Eq.~(\ref{eq:rescale}), pointing to sizable effects on carrier transport due to local disorder which exceed the mere mass rescaling.

Lastly, an important correction comes from taking into consideration the effect of electron-phonon coupling on the carrier effective masses.~We have calculated the band gap and mass renormalization at 300~K through the SDM~\cite{Zacharias2020}, by employing the ZG configuration with its antithetic pair~\cite{Zacharias2023} in a 6$\times$6$\times$6 supercell of the symmetric unit cell.~The ZG displacements have been generated by employing, accordingly, the 300~K SSCHA dynamical matrix.~This yields a band-gap decrease of 165~meV compared to the clamped-ion value in the primitive cell.~This value is in excellent agreement with the renormalization reported in Ref.~\cite{Zacharias2023}.~We obtain $m^*_{\text{el-ph}}=2.21\,m_0$, in agreement with the QP curvature computed by a cumulant Green's function approach~\cite{Zhou2019}.

Accounting now for the 300~K ZG displacements starting from the locally disordered structure, allows to combine both the effect of disorder and electron-phonon coupling.~The renormalization of the electronic structure is even more pronounced, with a band gap closing of 236~meV with respect to the disordered structure and an overall increase of 74~meV with respect to the symmetric unit cell.~The mass renormalization amounts to $m^*_{\text{dis+el-ph}}=2.89\,m_0$, that is $1.38\,m_e$, in agreement with the experimental data~\cite{Wang2016,Mechelen2008}.~The electron-phonon renormalization of both the band gap and effective mass are thus stronger on the disordered structure rather than the symmetric one.~Importantly, by taking into account the combined effect of both local disorder and electron-phonon coupling on the effective mass, as well as the dielectric rescaling, we obtain a more reasonable mobility overestimation of 290\% in SrTiO$_3$ with respect to the experimental value at 300~K.

\setlength\tabcolsep{0.4em}
\begin{table}[!ht]
\captionsetup[table]{skip=0pt}
 \captionof{table}{Band-gap renormalization ($\Delta E_\textup{gap}$), electron effective mass enhancement ($m^*/m_0$) and mobility overestimation at 300~K with respect to the experimental value, after dielectric rescaling. $m_0$ is the electron effective mass obtained in the unit cell (sym). The other calculations correspond to the inclusion of local disorder (dis), electron-phonon coupling (el-ph), and the combination of both effects (dis+el-ph).}
\centering
\begin{ruledtabular}
\begin{tabular}{c|c|ccc}
\mbox{}&\mbox{}&\mbox{$\Delta E_\textup{gap}$ (meV)}&\mbox{$m^*/m_0$}&\mbox{$\mu^*/\mu^\textup{exp}$} \\
\hline
\multirow{4}{*}{SrTiO$_3$}&\mbox{sym}&\mbox{-}&\mbox{1}&\mbox{838\%} \\
&\mbox{dis}&\mbox{310}&\mbox{1.22}&\mbox{686\%} \\
&\mbox{el-ph}&\mbox{-165}&\mbox{2.21}&\mbox{378\%} \\
&\mbox{dis+el-ph}&\mbox{74}&\mbox{2.89}&\mbox{290\%} \\
\hline
\multirow{2}{*}{KTaO$_3$}&\mbox{sym}&\mbox{-}&\mbox{1}&\mbox{345\%} \\
&\mbox{el-ph}&\mbox{-193}&\mbox{1.39}&\mbox{248\%} \\
\end{tabular}
\end{ruledtabular}
\label{tab:mass_enhancements}
\end{table}

\subsection{KTaO$_3$}
\label{sec:KTO}

\begin{figure}[b!]
        \includegraphics[width=0.5\textwidth, trim={0.4cm 0cm 0cm 0.45cm},clip]{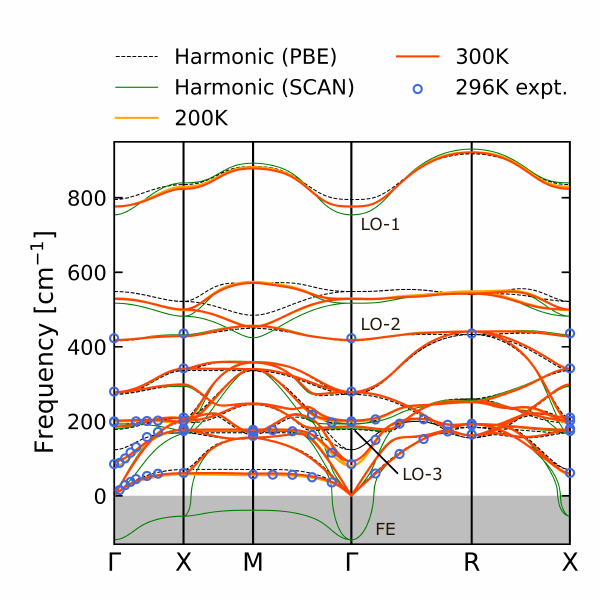}
        \vspace{-0.5cm}
        \caption{\label{fig:KTO_phonons}Stable harmonic phonon dispersion of KTaO$_3$ obtained using PBE (dashed black lines), unstable harmonic dispersion through the SCAN-based MLFF (solid green lines) and the temperature-dependent anharmonic dispersion using the SCAN-based MLFF in SSCHA (solid orange and red lines).~Experimental data at 296~K are also shown~\cite{Perry1967}.}
\end{figure}

\begin{figure}[t]
         \includegraphics[width=0.5\textwidth, trim={0.4cm 0.1cm 0.1cm 0.3cm},clip]{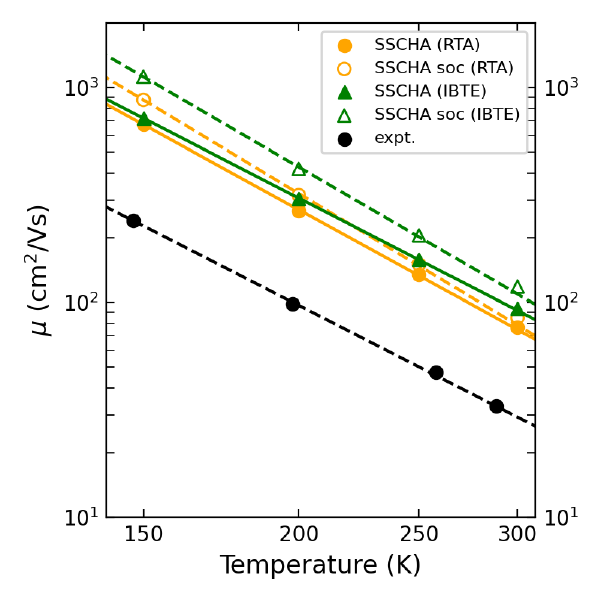}
        \vspace{-0.5cm}
        \caption{\label{fig:KTO_mobility}Electron mobility calculations in KTaO$_3$ using SSCHA temperature-dependent dynamical matrices at RTA (orange filled circles) and IBTE (green filled triangles) level. The same calculations are repeated including SOC (empty symbols).~Both the IBTE and the experimental mobilities \cite{Wemple1965} are fitted using Eq.~(\ref{eq:power_law}).~All the computed mobilities are rescaled by Eq.~(\ref{eq:rescale}).}
\end{figure}

We start by presenting the calculated temperature-dependent anharmonic phonon dispersions of KTaO$_3$ in Fig.~\ref{fig:KTO_phonons}, in excellent agreement with inelastic neutron scattering data~\cite{Perry1967}. The incipient ferroelectric instability is highly dependent on the exchange-correlation functional adopted.~The harmonic phonon dispersion in Fig.~\ref{fig:KTO_phonons} calculated by employing the PBE functional predicts a stable $\Gamma$-point FE mode.~The one obtained using the SCAN functional on a 4$\times$4$\times$4 supercell (commensurate with the q-grid employed for PBE), instead, is strongly unstable for the FE distortion, but stability is recovered by including the zero-point quantum energy and the anharmonic phonon effects through the MLFF+SSCHA scheme, as shown in Fig.~\ref{fig:KTO_phonons}. The temperature evolution of the soft FE mode, shown in SM Fig.~S6~\cite{SM}, is in excellent agreement with the experimental data. 
%thus improving on the previous results thanks to the use of a wider 4$\times$4$\times$4 supercell.

As in SrTiO$_3$, the temperature-insensitive LO-1, LO-2 and LO-3 modes are the main scattering channels, and a temperature-dependent contribution comes from the FE mode (see SM Fig.~1(c)-(d)~\cite{SM}).~By fitting the experimental data~\cite{Wemple1965} in Fig.~\ref{fig:KTO_mobility} with the power law of Eq.~(\ref{eq:power_law}), we find $\gamma=2.96$, similar to the temperature behavior of SrTiO$_3$.~The mobility calculated using the IBTE is in agreement with $\gamma=2.97$, while the RTA follows instead $\gamma=3.17$.~As in SrTiO$_3$, fixing the 150~K dynamical matrices leads to an underestimation of the electron mobility at room temperature, since the hardening of the FE mode is neglected (see SM Fig.~S8~\cite{SM}).

As for SrTiO$_3$ we analyze possible reasons for the mobility overestimation with respect to the experimental data.~The SOC-induced splitting of 314~meV near the CBM introduces only slight changes in the electron effective masses, as reported in SM Table S\rom{1}~\cite{SM}.~Nonetheless, the mobility is enhanced quite dramatically in comparison to the non-relativistic calculations in Fig.~\ref{fig:KTO_mobility}, while the power law dependence changes to $\gamma=3.36$ for IBTE and $\gamma=3.48$ for RTA. This mobility enhancement induced by SOC is thus to be attributed to band splitting in the 5$d$ manifold of the heavier Ta ion. 

The computed dielectric constant of~$\SI{5.07}{}$ slightly overestimates the experimental value of 4.6~\cite{Fujii1976}.~Through the dielectric rescaling in Eq.~(\ref{eq:rescale}) the mobility is reduced by 18\%, in analogy with SrTiO$_3$. Thus, at 300~K the mobility is overestimated by 345\% with respect to the experiment, a much lower value than in the case of SrTiO$_3$ (see the results summarized in Table~\ref{tab:mass_enhancements}).

We note that no local disorder has been found for the ground state of KTaO$_3$ using DFT-PBE in a 2$\times$2$\times$2 supercell.~As for SrTiO$_3$, we have computed the band gap renormalization and electron effective mass enhancement due to the electron-phonon interaction using the SDM.~Our calculations at 300~K yield a band-gap renormalization of $-0.193$~meV, which is similar to the one obtained for SrTiO$_3$.~Importantly, and at variance with this result, we obtain~$m^*_{\text{el-ph}}=0.57\,m_e$, a factor 1.4 greater than $m_0=0.41\,m_{e}$ at clamped nuclei. Hence, the total mass renormalization in KTaO$_3$ is much smaller than the one obtained for the locally disordered SrTiO$_3$. After accounting for dielectric screening and mobility rescaling due to effective mass renormalization, we obtain a mobility overestimation of 248\% with respect to the experimental value at 300~K, in line with the analogous estimate of 290\% for SrTiO$_3$.

\subsection{Other possible scattering mechanisms}
\label{sec:Other scattering mechanisms}
In our calculations, we did not account for additional phenomena that are expected to have a negligible impact on the computed mobility in SrTiO$_3$ and KTaO$_3$.~As regards the experimental samples integrity, we compare our results with measurements performed on single crystals of SrTiO$_3$ (grown by hybrid molecular beam epitaxy)~\cite{Cain2013} and KTaO$_3$ (grown by modified Kyropoulos method)~\cite{Wemple1965}. Consequently, we do not expect any carrier scattering due to the presence of grain boundaries.

Furthermore, the experimental samples at varying impurity density, and thus doping concentrations, show the same mobility values in the 150-300~K temperature range. This implies that impurity scattering is negligible, and that electron-electron scattering, in principle not negligible in heavily doped semiconductors~\cite{giustino2016}, is of minor importance in this case.

Our computed Hall mobility is only less than 3\% higher than the computed drift mobility for the whole temperature range, at both the RTA and IBTE level. This result justifies the use of the drift mobility in our analysis, and rules out the case of a different carrier transport behavior under an applied magnetic field in SrTiO$_3$ and KTaO$_3$.

The interpolation scheme adopted includes only the dipole-dipole correction to the electron-phonon matrix elements~\cite{Verdi2015}.~Recently, the quadrupole-quadrupole correction has been shown to substantially affect the mobility~\cite{Brunin2020b,brunin2020c,Jhalani2020}, mostly in the regime of strong acoustic modes scattering~\cite{ponce2021} rather than LO-dominated scattering. In the case of cubic SrTiO$_3$ and KTaO$_3$, the quadrupole tensor is null due to the $Pm3m$ space-group symmetry, but could have an impact for the locally disordered structure of global $Pm3m$ symmetry of SrTiO$_3$.

Lastly, the steeper slope in our computed mobility for the locally disordered SrTiO$_3$ with respect to the experiment may also indicate the presence of AFD and FE scattering channels beyond the 1-electron 1-phonon interaction.~Scattering with these modes may be stronger at lower temperatures and close to the AFD phase transition at 105~K in SrTiO$_3$. 

\section{CONCLUSION}
\label{sec:conclusions}

The temperature dependence of the electron mobility in SrTiO$_3$ and KTaO$_3$ in the phonon-limited regime has been recovered by the inclusion of the temperature-dependent anharmonic phonon dispersion in the Boltzmann transport equation.~The discrepancy with respect to the experiments is mitigated by (i) dielectric screening rescaling of the electron-phonon matrix elements, (ii) the inclusion of the electron eigenvalues associated to the supercell locally disordered structure in SrTiO$_3$, and (iii) accounting for temperature-dependent effective mass enhancements due to the electron-phonon interaction.~These contributions provide additional mechanisms to the previous explanation that considered the renormalization of the electron quasiparticle peaks and the presence of satellites due to the electron-phonon interaction~\cite{Zhou2019}.~The calculated mobilities in KTaO$_3$ are less overestimated as compared to the experiment due to the negligible presence of local disorder and lower electron-phonon effective mass enhancement.~Including the complete electronic dispersion in the BTE calculations shows that local disorder decreases the mobility substantially more than a simple mass rescaling.~Moreover, the sharper decline in mobility relative to experimental observations may hint at an electron-phonon scattering of higher order mediated by the  AFD and FE modes.

We hope this work will inspire further explorations into the impact of lattice anharmonicity, local disorder, and electron-phonon renormalization on the mobility and effective mass of charge carriers in systems with complex lattice dynamics, such as halide perovskites.

\section{ACKNOWLEDGEMENTS}

This work was supported by the Austrian Science Fund (FWF) projects I 4506 (FWO-FWF joint project) and SFB TACO (0.55776/F81).
~CV acknowledges support from the Australian Research Council (DE220101147).~M.Z.~was funded by the European Union (project ULTRA-2DPK / HORIZON-MSCA-2022-PF-01 / Grant Agreement No.~101106654).~Views and opinions expressed are however those of the authors only and do not necessarily reflect those of the European Union or the European Commission.~Neither the European Union nor the granting authority can be held responsible for them..~F.G. was supported by the Robert A. Welch Foundation under Award No. F-2139-20230405~The computational results presented have been achieved using the Vienna Scientific Cluster (VSC) and Frontera and Lonestar clusters from The Texas Advanced Computing Center (TACC) at the University of Texas at Austin. LR desires to acknowledge Michael Wolloch for his help in the construction of a faster workflow that combines SSCHA and the MLFF in VASP.

%\appendix
%\section{Appendixes}
%\subsection{\label{app:subsec}A subsection in an appendix}

%\clearpage
%


\begin{thebibliography}{97}%
\makeatletter
\providecommand \@ifxundefined [1]{%
 \@ifx{#1\undefined}
}%
\providecommand \@ifnum [1]{%
 \ifnum #1\expandafter \@firstoftwo
 \else \expandafter \@secondoftwo
 \fi
}%
\providecommand \@ifx [1]{%
 \ifx #1\expandafter \@firstoftwo
 \else \expandafter \@secondoftwo
 \fi
}%
\providecommand \natexlab [1]{#1}%
\providecommand \enquote  [1]{``#1''}%
\providecommand \bibnamefont  [1]{#1}%
\providecommand \bibfnamefont [1]{#1}%
\providecommand \citenamefont [1]{#1}%
\providecommand \href@noop [0]{\@secondoftwo}%
\providecommand \href [0]{\begingroup \@sanitize@url \@href}%
\providecommand \@href[1]{\@@startlink{#1}\@@href}%
\providecommand \@@href[1]{\endgroup#1\@@endlink}%
\providecommand \@sanitize@url [0]{\catcode `\\12\catcode `\$12\catcode `\&12\catcode `\#12\catcode `\^12\catcode `\_12\catcode `\%12\relax}%
\providecommand \@@startlink[1]{}%
\providecommand \@@endlink[0]{}%
\providecommand \url  [0]{\begingroup\@sanitize@url \@url }%
\providecommand \@url [1]{\endgroup\@href {#1}{\urlprefix }}%
\providecommand \urlprefix  [0]{URL }%
\providecommand \Eprint [0]{\href }%
\providecommand \doibase [0]{https://doi.org/}%
\providecommand \selectlanguage [0]{\@gobble}%
\providecommand \bibinfo  [0]{\@secondoftwo}%
\providecommand \bibfield  [0]{\@secondoftwo}%
\providecommand \translation [1]{[#1]}%
\providecommand \BibitemOpen [0]{}%
\providecommand \bibitemStop [0]{}%
\providecommand \bibitemNoStop [0]{.\EOS\space}%
\providecommand \EOS [0]{\spacefactor3000\relax}%
\providecommand \BibitemShut  [1]{\csname bibitem#1\endcsname}%
\let\auto@bib@innerbib\@empty
%</preamble>
\bibitem [{\citenamefont {Bednorz}\ and\ \citenamefont {M\"uller}(1988)}]{Bednorz1988}%
  \BibitemOpen
  \bibfield  {author} {\bibinfo {author} {\bibfnamefont {J.~G.}\ \bibnamefont {Bednorz}}\ and\ \bibinfo {author} {\bibfnamefont {K.~A.}\ \bibnamefont {M\"uller}},\ }\href {https://doi.org/10.1103/RevModPhys.60.585} {\bibfield  {journal} {\bibinfo  {journal} {Rev. Mod. Phys.}\ }\textbf {\bibinfo {volume} {60}},\ \bibinfo {pages} {585} (\bibinfo {year} {1988})}\BibitemShut {NoStop}%
\bibitem [{\citenamefont {Reyren}\ \emph {et~al.}(2007)\citenamefont {Reyren}, \citenamefont {Thiel}, \citenamefont {Caviglia}, \citenamefont {Kourkoutis}, \citenamefont {Hammerl}, \citenamefont {Richter}, \citenamefont {Schneider}, \citenamefont {Kopp}, \citenamefont {Rüetschi}, \citenamefont {Jaccard}, \citenamefont {Gabay}, \citenamefont {Muller}, \citenamefont {Triscone},\ and\ \citenamefont {Mannhart}}]{Reyren2007}%
  \BibitemOpen
  \bibfield  {author} {\bibinfo {author} {\bibfnamefont {N.}~\bibnamefont {Reyren}}, \bibinfo {author} {\bibfnamefont {S.}~\bibnamefont {Thiel}}, \bibinfo {author} {\bibfnamefont {A.~D.}\ \bibnamefont {Caviglia}}, \bibinfo {author} {\bibfnamefont {L.~F.}\ \bibnamefont {Kourkoutis}}, \bibinfo {author} {\bibfnamefont {G.}~\bibnamefont {Hammerl}}, \bibinfo {author} {\bibfnamefont {C.}~\bibnamefont {Richter}}, \bibinfo {author} {\bibfnamefont {C.~W.}\ \bibnamefont {Schneider}}, \bibinfo {author} {\bibfnamefont {T.}~\bibnamefont {Kopp}}, \bibinfo {author} {\bibfnamefont {A.~S.}\ \bibnamefont {Rüetschi}}, \bibinfo {author} {\bibfnamefont {D.}~\bibnamefont {Jaccard}}, \bibinfo {author} {\bibfnamefont {M.}~\bibnamefont {Gabay}}, \bibinfo {author} {\bibfnamefont {D.~A.}\ \bibnamefont {Muller}}, \bibinfo {author} {\bibfnamefont {J.~M.}\ \bibnamefont {Triscone}},\ and\ \bibinfo {author} {\bibfnamefont {J.}~\bibnamefont {Mannhart}},\ }\href {https://doi.org/10.1126/SCIENCE.1146006} {\bibfield  {journal} {\bibinfo
  {journal} {Science}\ }\textbf {\bibinfo {volume} {317}},\ \bibinfo {pages} {1196} (\bibinfo {year} {2007})}\BibitemShut {NoStop}%
\bibitem [{\citenamefont {Jin}\ \emph {et~al.}(1994)\citenamefont {Jin}, \citenamefont {Tiefel}, \citenamefont {McCormack}, \citenamefont {Fastnacht}, \citenamefont {Ramesh},\ and\ \citenamefont {Chen}}]{Jin1994}%
  \BibitemOpen
  \bibfield  {author} {\bibinfo {author} {\bibfnamefont {S.}~\bibnamefont {Jin}}, \bibinfo {author} {\bibfnamefont {T.~H.}\ \bibnamefont {Tiefel}}, \bibinfo {author} {\bibfnamefont {M.}~\bibnamefont {McCormack}}, \bibinfo {author} {\bibfnamefont {R.~A.}\ \bibnamefont {Fastnacht}}, \bibinfo {author} {\bibfnamefont {R.}~\bibnamefont {Ramesh}},\ and\ \bibinfo {author} {\bibfnamefont {L.~H.}\ \bibnamefont {Chen}},\ }\href {https://doi.org/10.1126/science.264.5157.413} {\bibfield  {journal} {\bibinfo  {journal} {Science}\ }\textbf {\bibinfo {volume} {264}},\ \bibinfo {pages} {413} (\bibinfo {year} {1994})}\BibitemShut {NoStop}%
\bibitem [{\citenamefont {Cohen}(1992)}]{Cohen1992}%
  \BibitemOpen
  \bibfield  {author} {\bibinfo {author} {\bibfnamefont {R.~E.}\ \bibnamefont {Cohen}},\ }\href {https://doi.org/10.1038/358136A0} {\bibfield  {journal} {\bibinfo  {journal} {Nature}\ }\textbf {\bibinfo {volume} {358}},\ \bibinfo {pages} {136} (\bibinfo {year} {1992})}\BibitemShut {NoStop}%
\bibitem [{\citenamefont {K.F.~Wang}\ and\ \citenamefont {Ren}(2009)}]{Wang2009}%
  \BibitemOpen
  \bibfield  {author} {\bibinfo {author} {\bibfnamefont {J.-M.~L.}\ \bibnamefont {K.F.~Wang}}\ and\ \bibinfo {author} {\bibfnamefont {Z.}~\bibnamefont {Ren}},\ }\href {https://doi.org/10.1080/00018730902920554} {\bibfield  {journal} {\bibinfo  {journal} {Adv. Phys.}\ }\textbf {\bibinfo {volume} {58}},\ \bibinfo {pages} {321} (\bibinfo {year} {2009})}\BibitemShut {NoStop}%
\bibitem [{\citenamefont {Narayan}\ \emph {et~al.}(2018)\citenamefont {Narayan}, \citenamefont {Cano}, \citenamefont {Balatsky},\ and\ \citenamefont {Spaldin}}]{Narayan2018}%
  \BibitemOpen
  \bibfield  {author} {\bibinfo {author} {\bibfnamefont {A.}~\bibnamefont {Narayan}}, \bibinfo {author} {\bibfnamefont {A.}~\bibnamefont {Cano}}, \bibinfo {author} {\bibfnamefont {A.~V.}\ \bibnamefont {Balatsky}},\ and\ \bibinfo {author} {\bibfnamefont {N.~A.}\ \bibnamefont {Spaldin}},\ }\href {https://doi.org/10.1038/s41563-018-0255-6} {\bibfield  {journal} {\bibinfo  {journal} {Nat. Mat.}\ }\textbf {\bibinfo {volume} {18}},\ \bibinfo {pages} {223} (\bibinfo {year} {2018})}\BibitemShut {NoStop}%
\bibitem [{\citenamefont {Li}\ \emph {et~al.}(2018)\citenamefont {Li}, \citenamefont {Lin}, \citenamefont {Chen}, \citenamefont {Cheng}, \citenamefont {Wang}, \citenamefont {Li}, \citenamefont {Xu}, \citenamefont {Huang}, \citenamefont {Liao}, \citenamefont {Chen}, \citenamefont {Shrout},\ and\ \citenamefont {Zhang}}]{Li2018}%
  \BibitemOpen
  \bibfield  {author} {\bibinfo {author} {\bibfnamefont {F.}~\bibnamefont {Li}}, \bibinfo {author} {\bibfnamefont {D.}~\bibnamefont {Lin}}, \bibinfo {author} {\bibfnamefont {Z.}~\bibnamefont {Chen}}, \bibinfo {author} {\bibfnamefont {Z.}~\bibnamefont {Cheng}}, \bibinfo {author} {\bibfnamefont {J.}~\bibnamefont {Wang}}, \bibinfo {author} {\bibfnamefont {C.}~\bibnamefont {Li}}, \bibinfo {author} {\bibfnamefont {Z.}~\bibnamefont {Xu}}, \bibinfo {author} {\bibfnamefont {Q.}~\bibnamefont {Huang}}, \bibinfo {author} {\bibfnamefont {X.}~\bibnamefont {Liao}}, \bibinfo {author} {\bibfnamefont {L.~Q.}\ \bibnamefont {Chen}}, \bibinfo {author} {\bibfnamefont {T.~R.}\ \bibnamefont {Shrout}},\ and\ \bibinfo {author} {\bibfnamefont {S.}~\bibnamefont {Zhang}},\ }\href {https://doi.org/10.1038/S41563-018-0034-4} {\bibfield  {journal} {\bibinfo  {journal} {Nat. Mater.}\ }\textbf {\bibinfo {volume} {17}},\ \bibinfo {pages} {349} (\bibinfo {year} {2018})}\BibitemShut {NoStop}%
\bibitem [{\citenamefont {Imada}\ \emph {et~al.}(1998)\citenamefont {Imada}, \citenamefont {Fujimori},\ and\ \citenamefont {Tokura}}]{Imada1998}%
  \BibitemOpen
  \bibfield  {author} {\bibinfo {author} {\bibfnamefont {M.}~\bibnamefont {Imada}}, \bibinfo {author} {\bibfnamefont {A.}~\bibnamefont {Fujimori}},\ and\ \bibinfo {author} {\bibfnamefont {Y.}~\bibnamefont {Tokura}},\ }\href {https://doi.org/10.1103/RevModPhys.70.1039} {\bibfield  {journal} {\bibinfo  {journal} {Rev. Mod. Phys.}\ }\textbf {\bibinfo {volume} {70}},\ \bibinfo {pages} {1039} (\bibinfo {year} {1998})}\BibitemShut {NoStop}%
\bibitem [{\citenamefont {He}\ and\ \citenamefont {Franchini}(2012)}]{He2012}%
  \BibitemOpen
  \bibfield  {author} {\bibinfo {author} {\bibfnamefont {J.}~\bibnamefont {He}}\ and\ \bibinfo {author} {\bibfnamefont {C.}~\bibnamefont {Franchini}},\ }\href {https://doi.org/10.1103/PhysRevB.86.235117} {\bibfield  {journal} {\bibinfo  {journal} {Phys. Rev. B}\ }\textbf {\bibinfo {volume} {86}},\ \bibinfo {pages} {235117} (\bibinfo {year} {2012})}\BibitemShut {NoStop}%
\bibitem [{\citenamefont {Erg\"onenc}\ \emph {et~al.}(2018)\citenamefont {Erg\"onenc}, \citenamefont {Kim}, \citenamefont {Liu}, \citenamefont {Kresse},\ and\ \citenamefont {Franchini}}]{Ergonenc2018}%
  \BibitemOpen
  \bibfield  {author} {\bibinfo {author} {\bibfnamefont {Z.}~\bibnamefont {Erg\"onenc}}, \bibinfo {author} {\bibfnamefont {B.}~\bibnamefont {Kim}}, \bibinfo {author} {\bibfnamefont {P.}~\bibnamefont {Liu}}, \bibinfo {author} {\bibfnamefont {G.}~\bibnamefont {Kresse}},\ and\ \bibinfo {author} {\bibfnamefont {C.}~\bibnamefont {Franchini}},\ }\href {https://doi.org/10.1103/PhysRevMaterials.2.024601} {\bibfield  {journal} {\bibinfo  {journal} {Phys. Rev. Mater.}\ }\textbf {\bibinfo {volume} {2}},\ \bibinfo {pages} {024601} (\bibinfo {year} {2018})}\BibitemShut {NoStop}%
\bibitem [{\citenamefont {Varrassi}\ \emph {et~al.}(2021)\citenamefont {Varrassi}, \citenamefont {Liu}, \citenamefont {Yavas}, \citenamefont {Bokdam}, \citenamefont {Kresse},\ and\ \citenamefont {Franchini}}]{Varrassi2021}%
  \BibitemOpen
  \bibfield  {author} {\bibinfo {author} {\bibfnamefont {L.}~\bibnamefont {Varrassi}}, \bibinfo {author} {\bibfnamefont {P.}~\bibnamefont {Liu}}, \bibinfo {author} {\bibfnamefont {Z.~E.}\ \bibnamefont {Yavas}}, \bibinfo {author} {\bibfnamefont {M.}~\bibnamefont {Bokdam}}, \bibinfo {author} {\bibfnamefont {G.}~\bibnamefont {Kresse}},\ and\ \bibinfo {author} {\bibfnamefont {C.}~\bibnamefont {Franchini}},\ }\href {https://doi.org/10.1103/PhysRevMaterials.5.074601} {\bibfield  {journal} {\bibinfo  {journal} {Phys. Rev. Mater.}\ }\textbf {\bibinfo {volume} {5}},\ \bibinfo {pages} {074601} (\bibinfo {year} {2021})}\BibitemShut {NoStop}%
\bibitem [{\citenamefont {Liu}\ \emph {et~al.}(2024)\citenamefont {Liu}, \citenamefont {Celiberti}, \citenamefont {Decker}, \citenamefont {Ruotsalainen}, \citenamefont {Siewierska}, \citenamefont {Kusch}, \citenamefont {Wang}, \citenamefont {Kim}, \citenamefont {Olaniyan}, \citenamefont {Di~Castro}, \citenamefont {Tomiyasu}, \citenamefont {van~der Minne}, \citenamefont {Birkh{\"o}lzer}, \citenamefont {Kiens}, \citenamefont {van~den Bosch}, \citenamefont {Patil}, \citenamefont {Baeumer}, \citenamefont {Koster}, \citenamefont {Lazemi}, \citenamefont {de~Groot}, \citenamefont {Dubourdieu}, \citenamefont {Franchini},\ and\ \citenamefont {F{\"o}hlisch}}]{Liu2024}%
  \BibitemOpen
  \bibfield  {author} {\bibinfo {author} {\bibfnamefont {C.-Y.}\ \bibnamefont {Liu}}, \bibinfo {author} {\bibfnamefont {L.}~\bibnamefont {Celiberti}}, \bibinfo {author} {\bibfnamefont {R.}~\bibnamefont {Decker}}, \bibinfo {author} {\bibfnamefont {K.}~\bibnamefont {Ruotsalainen}}, \bibinfo {author} {\bibfnamefont {K.}~\bibnamefont {Siewierska}}, \bibinfo {author} {\bibfnamefont {M.}~\bibnamefont {Kusch}}, \bibinfo {author} {\bibfnamefont {R.-P.}\ \bibnamefont {Wang}}, \bibinfo {author} {\bibfnamefont {D.~J.}\ \bibnamefont {Kim}}, \bibinfo {author} {\bibfnamefont {I.~I.}\ \bibnamefont {Olaniyan}}, \bibinfo {author} {\bibfnamefont {D.}~\bibnamefont {Di~Castro}}, \bibinfo {author} {\bibfnamefont {K.}~\bibnamefont {Tomiyasu}}, \bibinfo {author} {\bibfnamefont {E.}~\bibnamefont {van~der Minne}}, \bibinfo {author} {\bibfnamefont {Y.~A.}\ \bibnamefont {Birkh{\"o}lzer}}, \bibinfo {author} {\bibfnamefont {E.~M.}\ \bibnamefont {Kiens}}, \bibinfo {author} {\bibfnamefont {I.~C.~G.}\ \bibnamefont {van~den Bosch}}, \bibinfo
  {author} {\bibfnamefont {K.~N.}\ \bibnamefont {Patil}}, \bibinfo {author} {\bibfnamefont {C.}~\bibnamefont {Baeumer}}, \bibinfo {author} {\bibfnamefont {G.}~\bibnamefont {Koster}}, \bibinfo {author} {\bibfnamefont {M.}~\bibnamefont {Lazemi}}, \bibinfo {author} {\bibfnamefont {F.~M.~F.}\ \bibnamefont {de~Groot}}, \bibinfo {author} {\bibfnamefont {C.}~\bibnamefont {Dubourdieu}}, \bibinfo {author} {\bibfnamefont {C.}~\bibnamefont {Franchini}},\ and\ \bibinfo {author} {\bibfnamefont {A.}~\bibnamefont {F{\"o}hlisch}},\ }\href {https://doi.org/10.1038/s42005-024-01642-5} {\bibfield  {journal} {\bibinfo  {journal} {Commun. Phys.}\ }\textbf {\bibinfo {volume} {7}},\ \bibinfo {pages} {156} (\bibinfo {year} {2024})}\BibitemShut {NoStop}%
\bibitem [{\citenamefont {Fatih~Dogan}\ and\ \citenamefont {Peña}(2015)}]{Dogan2015}%
  \BibitemOpen
  \bibfield  {author} {\bibinfo {author} {\bibfnamefont {M.~G.-V.}\ \bibnamefont {Fatih~Dogan}, \bibfnamefont {Hong~Lin}}\ and\ \bibinfo {author} {\bibfnamefont {O.}~\bibnamefont {Peña}},\ }\href {https://doi.org/10.1088/1468-6996/16/2/020301} {\bibfield  {journal} {\bibinfo  {journal} {Sci. Technol. Adv. Mater.}\ }\textbf {\bibinfo {volume} {16}},\ \bibinfo {pages} {020301} (\bibinfo {year} {2015})}\BibitemShut {NoStop}%
\bibitem [{\citenamefont {Sawa}(2008)}]{Sawa2008}%
  \BibitemOpen
  \bibfield  {author} {\bibinfo {author} {\bibfnamefont {A.}~\bibnamefont {Sawa}},\ }\href {https://doi.org/10.1016/S1369-7021(08)70119-6} {\bibfield  {journal} {\bibinfo  {journal} {Mater. Today}\ }\textbf {\bibinfo {volume} {11}},\ \bibinfo {pages} {28} (\bibinfo {year} {2008})}\BibitemShut {NoStop}%
\bibitem [{\citenamefont {Sun}\ \emph {et~al.}(2021)\citenamefont {Sun}, \citenamefont {Zhou}, \citenamefont {Sun}, \citenamefont {Zhao}, \citenamefont {Chen}, \citenamefont {Yang}, \citenamefont {Zhao},\ and\ \citenamefont {Song}}]{Sun2021}%
  \BibitemOpen
  \bibfield  {author} {\bibinfo {author} {\bibfnamefont {B.}~\bibnamefont {Sun}}, \bibinfo {author} {\bibfnamefont {G.}~\bibnamefont {Zhou}}, \bibinfo {author} {\bibfnamefont {L.}~\bibnamefont {Sun}}, \bibinfo {author} {\bibfnamefont {H.}~\bibnamefont {Zhao}}, \bibinfo {author} {\bibfnamefont {Y.}~\bibnamefont {Chen}}, \bibinfo {author} {\bibfnamefont {F.}~\bibnamefont {Yang}}, \bibinfo {author} {\bibfnamefont {Y.}~\bibnamefont {Zhao}},\ and\ \bibinfo {author} {\bibfnamefont {Q.}~\bibnamefont {Song}},\ }\href {https://doi.org/10.1039/D1NH00292A} {\bibfield  {journal} {\bibinfo  {journal} {Nanoscale Horiz}\ }\textbf {\bibinfo {volume} {6}},\ \bibinfo {pages} {939} (\bibinfo {year} {2021})}\BibitemShut {NoStop}%
\bibitem [{\citenamefont {Kim}\ \emph {et~al.}(2014)\citenamefont {Kim}, \citenamefont {Im}, \citenamefont {Freeman}, \citenamefont {Ihm},\ and\ \citenamefont {Jin}}]{Kim2014}%
  \BibitemOpen
  \bibfield  {author} {\bibinfo {author} {\bibfnamefont {M.}~\bibnamefont {Kim}}, \bibinfo {author} {\bibfnamefont {J.}~\bibnamefont {Im}}, \bibinfo {author} {\bibfnamefont {A.~J.}\ \bibnamefont {Freeman}}, \bibinfo {author} {\bibfnamefont {J.}~\bibnamefont {Ihm}},\ and\ \bibinfo {author} {\bibfnamefont {H.}~\bibnamefont {Jin}},\ }\href {https://doi.org/10.1073/PNAS.1405780111} {\bibfield  {journal} {\bibinfo  {journal} {Proc. Natl. Acad. Sci. U. S. A.}\ }\textbf {\bibinfo {volume} {111}},\ \bibinfo {pages} {6900} (\bibinfo {year} {2014})}\BibitemShut {NoStop}%
\bibitem [{\citenamefont {Suntivich}\ \emph {et~al.}(2011)\citenamefont {Suntivich}, \citenamefont {Gasteiger}, \citenamefont {Yabuuchi}, \citenamefont {Nakanishi}, \citenamefont {Goodenough},\ and\ \citenamefont {Shao-Horn}}]{Suntivich2011}%
  \BibitemOpen
  \bibfield  {author} {\bibinfo {author} {\bibfnamefont {J.}~\bibnamefont {Suntivich}}, \bibinfo {author} {\bibfnamefont {H.~A.}\ \bibnamefont {Gasteiger}}, \bibinfo {author} {\bibfnamefont {N.}~\bibnamefont {Yabuuchi}}, \bibinfo {author} {\bibfnamefont {H.}~\bibnamefont {Nakanishi}}, \bibinfo {author} {\bibfnamefont {J.~B.}\ \bibnamefont {Goodenough}},\ and\ \bibinfo {author} {\bibfnamefont {Y.}~\bibnamefont {Shao-Horn}},\ }\href {https://doi.org/10.1038/NCHEM.1069} {\bibfield  {journal} {\bibinfo  {journal} {Nat. Chem.}\ }\textbf {\bibinfo {volume} {3}},\ \bibinfo {pages} {546} (\bibinfo {year} {2011})}\BibitemShut {NoStop}%
\bibitem [{\citenamefont {Ponc\'e}\ \emph {et~al.}(2018)\citenamefont {Ponc\'e}, \citenamefont {Margine},\ and\ \citenamefont {Giustino}}]{ponce2018}%
  \BibitemOpen
  \bibfield  {author} {\bibinfo {author} {\bibfnamefont {S.}~\bibnamefont {Ponc\'e}}, \bibinfo {author} {\bibfnamefont {E.~R.}\ \bibnamefont {Margine}},\ and\ \bibinfo {author} {\bibfnamefont {F.}~\bibnamefont {Giustino}},\ }\href {https://doi.org/10.1103/PhysRevB.97.121201} {\bibfield  {journal} {\bibinfo  {journal} {Phys. Rev. B}\ }\textbf {\bibinfo {volume} {97}},\ \bibinfo {pages} {121201(R)} (\bibinfo {year} {2018})}\BibitemShut {NoStop}%
\bibitem [{\citenamefont {Protik}\ and\ \citenamefont {Broido}(2020)}]{Protik2020a}%
  \BibitemOpen
  \bibfield  {author} {\bibinfo {author} {\bibfnamefont {N.~H.}\ \bibnamefont {Protik}}\ and\ \bibinfo {author} {\bibfnamefont {D.~A.}\ \bibnamefont {Broido}},\ }\href {https://doi.org/10.1103/PhysRevB.101.075202} {\bibfield  {journal} {\bibinfo  {journal} {Phys. Rev. B}\ }\textbf {\bibinfo {volume} {101}},\ \bibinfo {pages} {075202} (\bibinfo {year} {2020})}\BibitemShut {NoStop}%
\bibitem [{\citenamefont {Park}\ \emph {et~al.}(2020)\citenamefont {Park}, \citenamefont {Zhou}, \citenamefont {Jhalani}, \citenamefont {Dreyer},\ and\ \citenamefont {Bernardi}}]{Park2020}%
  \BibitemOpen
  \bibfield  {author} {\bibinfo {author} {\bibfnamefont {J.}~\bibnamefont {Park}}, \bibinfo {author} {\bibfnamefont {J.-J.}\ \bibnamefont {Zhou}}, \bibinfo {author} {\bibfnamefont {V.~A.}\ \bibnamefont {Jhalani}}, \bibinfo {author} {\bibfnamefont {C.~E.}\ \bibnamefont {Dreyer}},\ and\ \bibinfo {author} {\bibfnamefont {M.}~\bibnamefont {Bernardi}},\ }\href {https://doi.org/10.1103/PhysRevB.102.125203} {\bibfield  {journal} {\bibinfo  {journal} {Phys. Rev. B}\ }\textbf {\bibinfo {volume} {102}},\ \bibinfo {pages} {125203} (\bibinfo {year} {2020})}\BibitemShut {NoStop}%
\bibitem [{\citenamefont {Zhou}\ \emph {et~al.}(2021)\citenamefont {Zhou}, \citenamefont {Park}, \citenamefont {Lu}, \citenamefont {Maliyov}, \citenamefont {Tong},\ and\ \citenamefont {Bernardi}}]{Zhou2021}%
  \BibitemOpen
  \bibfield  {author} {\bibinfo {author} {\bibfnamefont {J.~J.}\ \bibnamefont {Zhou}}, \bibinfo {author} {\bibfnamefont {J.}~\bibnamefont {Park}}, \bibinfo {author} {\bibfnamefont {I.~T.}\ \bibnamefont {Lu}}, \bibinfo {author} {\bibfnamefont {I.}~\bibnamefont {Maliyov}}, \bibinfo {author} {\bibfnamefont {X.}~\bibnamefont {Tong}},\ and\ \bibinfo {author} {\bibfnamefont {M.}~\bibnamefont {Bernardi}},\ }\href {https://doi.org/10.1016/J.CPC.2021.107970} {\bibfield  {journal} {\bibinfo  {journal} {Comput. Phys. Commun.}\ }\textbf {\bibinfo {volume} {264}},\ \bibinfo {pages} {107970} (\bibinfo {year} {2021})}\BibitemShut {NoStop}%
\bibitem [{\citenamefont {Gonze}\ \emph {et~al.}(2020)\citenamefont {Gonze}, \citenamefont {Amadon}, \citenamefont {Antonius}, \citenamefont {Arnardi}, \citenamefont {Baguet}, \citenamefont {Beuken}, \citenamefont {Bieder}, \citenamefont {Bottin}, \citenamefont {Bouchet}, \citenamefont {Bousquet}, \citenamefont {Brouwer}, \citenamefont {Bruneval}, \citenamefont {Brunin}, \citenamefont {Cavignac}, \citenamefont {Charraud}, \citenamefont {Chen}, \citenamefont {Côté}, \citenamefont {Cottenier}, \citenamefont {Denier}, \citenamefont {Geneste}, \citenamefont {Ghosez}, \citenamefont {Giantomassi}, \citenamefont {Gillet}, \citenamefont {Gingras}, \citenamefont {Hamann}, \citenamefont {Hautier}, \citenamefont {He}, \citenamefont {Helbig}, \citenamefont {Holzwarth}, \citenamefont {Jia}, \citenamefont {Jollet}, \citenamefont {Lafargue-Dit-Hauret}, \citenamefont {Lejaeghere}, \citenamefont {Marques}, \citenamefont {Martin}, \citenamefont {Martins}, \citenamefont {Miranda}, \citenamefont {Naccarato}, \citenamefont
  {Persson}, \citenamefont {Petretto}, \citenamefont {Planes}, \citenamefont {Pouillon}, \citenamefont {Prokhorenko}, \citenamefont {Ricci}, \citenamefont {Rignanese}, \citenamefont {Romero}, \citenamefont {Schmitt}, \citenamefont {Torrent}, \citenamefont {van Setten}, \citenamefont {Troeye}, \citenamefont {Verstraete}, \citenamefont {Zérah},\ and\ \citenamefont {Zwanziger}}]{Gonze2020}%
  \BibitemOpen
  \bibfield  {author} {\bibinfo {author} {\bibfnamefont {X.}~\bibnamefont {Gonze}}, \bibinfo {author} {\bibfnamefont {B.}~\bibnamefont {Amadon}}, \bibinfo {author} {\bibfnamefont {G.}~\bibnamefont {Antonius}}, \bibinfo {author} {\bibfnamefont {F.}~\bibnamefont {Arnardi}}, \bibinfo {author} {\bibfnamefont {L.}~\bibnamefont {Baguet}}, \bibinfo {author} {\bibfnamefont {J.~M.}\ \bibnamefont {Beuken}}, \bibinfo {author} {\bibfnamefont {J.}~\bibnamefont {Bieder}}, \bibinfo {author} {\bibfnamefont {F.}~\bibnamefont {Bottin}}, \bibinfo {author} {\bibfnamefont {J.}~\bibnamefont {Bouchet}}, \bibinfo {author} {\bibfnamefont {E.}~\bibnamefont {Bousquet}}, \bibinfo {author} {\bibfnamefont {N.}~\bibnamefont {Brouwer}}, \bibinfo {author} {\bibfnamefont {F.}~\bibnamefont {Bruneval}}, \bibinfo {author} {\bibfnamefont {G.}~\bibnamefont {Brunin}}, \bibinfo {author} {\bibfnamefont {T.}~\bibnamefont {Cavignac}}, \bibinfo {author} {\bibfnamefont {J.~B.}\ \bibnamefont {Charraud}}, \bibinfo {author} {\bibfnamefont {W.}~\bibnamefont
  {Chen}}, \bibinfo {author} {\bibfnamefont {M.}~\bibnamefont {Côté}}, \bibinfo {author} {\bibfnamefont {S.}~\bibnamefont {Cottenier}}, \bibinfo {author} {\bibfnamefont {J.}~\bibnamefont {Denier}}, \bibinfo {author} {\bibfnamefont {G.}~\bibnamefont {Geneste}}, \bibinfo {author} {\bibfnamefont {P.}~\bibnamefont {Ghosez}}, \bibinfo {author} {\bibfnamefont {M.}~\bibnamefont {Giantomassi}}, \bibinfo {author} {\bibfnamefont {Y.}~\bibnamefont {Gillet}}, \bibinfo {author} {\bibfnamefont {O.}~\bibnamefont {Gingras}}, \bibinfo {author} {\bibfnamefont {D.~R.}\ \bibnamefont {Hamann}}, \bibinfo {author} {\bibfnamefont {G.}~\bibnamefont {Hautier}}, \bibinfo {author} {\bibfnamefont {X.}~\bibnamefont {He}}, \bibinfo {author} {\bibfnamefont {N.}~\bibnamefont {Helbig}}, \bibinfo {author} {\bibfnamefont {N.}~\bibnamefont {Holzwarth}}, \bibinfo {author} {\bibfnamefont {Y.}~\bibnamefont {Jia}}, \bibinfo {author} {\bibfnamefont {F.}~\bibnamefont {Jollet}}, \bibinfo {author} {\bibfnamefont {W.}~\bibnamefont
  {Lafargue-Dit-Hauret}}, \bibinfo {author} {\bibfnamefont {K.}~\bibnamefont {Lejaeghere}}, \bibinfo {author} {\bibfnamefont {M.~A.}\ \bibnamefont {Marques}}, \bibinfo {author} {\bibfnamefont {A.}~\bibnamefont {Martin}}, \bibinfo {author} {\bibfnamefont {C.}~\bibnamefont {Martins}}, \bibinfo {author} {\bibfnamefont {H.~P.}\ \bibnamefont {Miranda}}, \bibinfo {author} {\bibfnamefont {F.}~\bibnamefont {Naccarato}}, \bibinfo {author} {\bibfnamefont {K.}~\bibnamefont {Persson}}, \bibinfo {author} {\bibfnamefont {G.}~\bibnamefont {Petretto}}, \bibinfo {author} {\bibfnamefont {V.}~\bibnamefont {Planes}}, \bibinfo {author} {\bibfnamefont {Y.}~\bibnamefont {Pouillon}}, \bibinfo {author} {\bibfnamefont {S.}~\bibnamefont {Prokhorenko}}, \bibinfo {author} {\bibfnamefont {F.}~\bibnamefont {Ricci}}, \bibinfo {author} {\bibfnamefont {G.~M.}\ \bibnamefont {Rignanese}}, \bibinfo {author} {\bibfnamefont {A.~H.}\ \bibnamefont {Romero}}, \bibinfo {author} {\bibfnamefont {M.~M.}\ \bibnamefont {Schmitt}}, \bibinfo {author}
  {\bibfnamefont {M.}~\bibnamefont {Torrent}}, \bibinfo {author} {\bibfnamefont {M.~J.}\ \bibnamefont {van Setten}}, \bibinfo {author} {\bibfnamefont {B.~V.}\ \bibnamefont {Troeye}}, \bibinfo {author} {\bibfnamefont {M.~J.}\ \bibnamefont {Verstraete}}, \bibinfo {author} {\bibfnamefont {G.}~\bibnamefont {Zérah}},\ and\ \bibinfo {author} {\bibfnamefont {J.~W.}\ \bibnamefont {Zwanziger}},\ }\href {https://doi.org/10.1016/J.CPC.2019.107042} {\bibfield  {journal} {\bibinfo  {journal} {Comput. Phys. Comm.}\ }\textbf {\bibinfo {volume} {248}},\ \bibinfo {pages} {107042} (\bibinfo {year} {2020})}\BibitemShut {NoStop}%
\bibitem [{\citenamefont {Restrepo}\ \emph {et~al.}(2009)\citenamefont {Restrepo}, \citenamefont {Varga},\ and\ \citenamefont {Pantelides}}]{Restrepo2009}%
  \BibitemOpen
  \bibfield  {author} {\bibinfo {author} {\bibfnamefont {O.~D.}\ \bibnamefont {Restrepo}}, \bibinfo {author} {\bibfnamefont {K.}~\bibnamefont {Varga}},\ and\ \bibinfo {author} {\bibfnamefont {S.~T.}\ \bibnamefont {Pantelides}},\ }\href {https://doi.org/10.1063/1.3147189} {\bibfield  {journal} {\bibinfo  {journal} {Appl. Phys. Lett.}\ }\textbf {\bibinfo {volume} {94}},\ \bibinfo {pages} {212103} (\bibinfo {year} {2009})}\BibitemShut {NoStop}%
\bibitem [{\citenamefont {Zhou}\ and\ \citenamefont {Bernardi}(2016)}]{Zhou2016}%
  \BibitemOpen
  \bibfield  {author} {\bibinfo {author} {\bibfnamefont {J.-J.}\ \bibnamefont {Zhou}}\ and\ \bibinfo {author} {\bibfnamefont {M.}~\bibnamefont {Bernardi}},\ }\href {https://doi.org/10.1103/PhysRevB.94.201201} {\bibfield  {journal} {\bibinfo  {journal} {Phys. Rev. B}\ }\textbf {\bibinfo {volume} {94}},\ \bibinfo {pages} {201201(R)} (\bibinfo {year} {2016})}\BibitemShut {NoStop}%
\bibitem [{\citenamefont {Zhao}\ \emph {et~al.}(2020{\natexlab{a}})\citenamefont {Zhao}, \citenamefont {Lian}, \citenamefont {Zeng}, \citenamefont {Dai}, \citenamefont {Meng},\ and\ \citenamefont {Ni}}]{Zhao2020b}%
  \BibitemOpen
  \bibfield  {author} {\bibinfo {author} {\bibfnamefont {Y.}~\bibnamefont {Zhao}}, \bibinfo {author} {\bibfnamefont {C.}~\bibnamefont {Lian}}, \bibinfo {author} {\bibfnamefont {S.}~\bibnamefont {Zeng}}, \bibinfo {author} {\bibfnamefont {Z.}~\bibnamefont {Dai}}, \bibinfo {author} {\bibfnamefont {S.}~\bibnamefont {Meng}},\ and\ \bibinfo {author} {\bibfnamefont {J.}~\bibnamefont {Ni}},\ }\href {https://doi.org/10.1103/PhysRevB.102.094314} {\bibfield  {journal} {\bibinfo  {journal} {Phys. Rev. B}\ }\textbf {\bibinfo {volume} {102}},\ \bibinfo {pages} {094314} (\bibinfo {year} {2020}{\natexlab{a}})}\BibitemShut {NoStop}%
\bibitem [{\citenamefont {Protik}\ and\ \citenamefont {Kozinsky}(2020)}]{Protik2020b}%
  \BibitemOpen
  \bibfield  {author} {\bibinfo {author} {\bibfnamefont {N.~H.}\ \bibnamefont {Protik}}\ and\ \bibinfo {author} {\bibfnamefont {B.}~\bibnamefont {Kozinsky}},\ }\href {https://doi.org/10.1103/PhysRevB.102.245202} {\bibfield  {journal} {\bibinfo  {journal} {Phys. Rev. B}\ }\textbf {\bibinfo {volume} {102}},\ \bibinfo {pages} {245202} (\bibinfo {year} {2020})}\BibitemShut {NoStop}%
\bibitem [{\citenamefont {Ponc\'e}\ \emph {et~al.}(2019{\natexlab{a}})\citenamefont {Ponc\'e}, \citenamefont {Jena},\ and\ \citenamefont {Giustino}}]{Ponce2019a}%
  \BibitemOpen
  \bibfield  {author} {\bibinfo {author} {\bibfnamefont {S.}~\bibnamefont {Ponc\'e}}, \bibinfo {author} {\bibfnamefont {D.}~\bibnamefont {Jena}},\ and\ \bibinfo {author} {\bibfnamefont {F.}~\bibnamefont {Giustino}},\ }\bibfield  {title} {\bibinfo {title} {Hole mobility of strained gan from first principles},\ }\href {https://doi.org/10.1103/PhysRevB.100.085204} {\bibfield  {journal} {\bibinfo  {journal} {Phys. Rev. B}\ }\textbf {\bibinfo {volume} {100}},\ \bibinfo {pages} {085204} (\bibinfo {year} {2019}{\natexlab{a}})}\BibitemShut {NoStop}%
\bibitem [{\citenamefont {Ponc\'e}\ \emph {et~al.}(2019{\natexlab{b}})\citenamefont {Ponc\'e}, \citenamefont {Jena},\ and\ \citenamefont {Giustino}}]{ponce2019b}%
  \BibitemOpen
  \bibfield  {author} {\bibinfo {author} {\bibfnamefont {S.}~\bibnamefont {Ponc\'e}}, \bibinfo {author} {\bibfnamefont {D.}~\bibnamefont {Jena}},\ and\ \bibinfo {author} {\bibfnamefont {F.}~\bibnamefont {Giustino}},\ }\href {https://doi.org/10.1103/PhysRevLett.123.096602} {\bibfield  {journal} {\bibinfo  {journal} {Phys. Rev. Lett.}\ }\textbf {\bibinfo {volume} {123}},\ \bibinfo {pages} {096602} (\bibinfo {year} {2019}{\natexlab{b}})}\BibitemShut {NoStop}%
\bibitem [{\citenamefont {Ponc\'e}\ and\ \citenamefont {Giustino}(2020)}]{ponce2020}%
  \BibitemOpen
  \bibfield  {author} {\bibinfo {author} {\bibfnamefont {S.}~\bibnamefont {Ponc\'e}}\ and\ \bibinfo {author} {\bibfnamefont {F.}~\bibnamefont {Giustino}},\ }\href {https://doi.org/10.1103/PhysRevResearch.2.033102} {\bibfield  {journal} {\bibinfo  {journal} {Phys. Rev. Res.}\ }\textbf {\bibinfo {volume} {2}},\ \bibinfo {pages} {033102} (\bibinfo {year} {2020})}\BibitemShut {NoStop}%
\bibitem [{\citenamefont {Ponc\'e}\ \emph {et~al.}(2020)\citenamefont {Ponc\'e}, \citenamefont {Li}, \citenamefont {Reichardt},\ and\ \citenamefont {Giustino}}]{ponce2020b}%
  \BibitemOpen
  \bibfield  {author} {\bibinfo {author} {\bibfnamefont {S.}~\bibnamefont {Ponc\'e}}, \bibinfo {author} {\bibfnamefont {W.}~\bibnamefont {Li}}, \bibinfo {author} {\bibfnamefont {S.}~\bibnamefont {Reichardt}},\ and\ \bibinfo {author} {\bibfnamefont {F.}~\bibnamefont {Giustino}},\ }\href {https://doi.org/10.1088/1361-6633/AB6A43} {\bibfield  {journal} {\bibinfo  {journal} {Rep. Prog. Phys.}\ }\textbf {\bibinfo {volume} {83}},\ \bibinfo {pages} {036501} (\bibinfo {year} {2020})}\BibitemShut {NoStop}%
\bibitem [{\citenamefont {Ponc\'e}\ \emph {et~al.}(2021)\citenamefont {Ponc\'e}, \citenamefont {Macheda}, \citenamefont {Margine}, \citenamefont {Marzari}, \citenamefont {Bonini},\ and\ \citenamefont {Giustino}}]{ponce2021}%
  \BibitemOpen
  \bibfield  {author} {\bibinfo {author} {\bibfnamefont {S.}~\bibnamefont {Ponc\'e}}, \bibinfo {author} {\bibfnamefont {F.}~\bibnamefont {Macheda}}, \bibinfo {author} {\bibfnamefont {E.~R.}\ \bibnamefont {Margine}}, \bibinfo {author} {\bibfnamefont {N.}~\bibnamefont {Marzari}}, \bibinfo {author} {\bibfnamefont {N.}~\bibnamefont {Bonini}},\ and\ \bibinfo {author} {\bibfnamefont {F.}~\bibnamefont {Giustino}},\ }\href {https://doi.org/10.1103/PHYSREVRESEARCH.3.043022} {\bibfield  {journal} {\bibinfo  {journal} {Phys. Rev. Res.}\ }\textbf {\bibinfo {volume} {3}},\ \bibinfo {pages} {043022} (\bibinfo {year} {2021})}\BibitemShut {NoStop}%
\bibitem [{\citenamefont {Macheda}\ and\ \citenamefont {Bonini}(2018)}]{Macheda2018}%
  \BibitemOpen
  \bibfield  {author} {\bibinfo {author} {\bibfnamefont {F.}~\bibnamefont {Macheda}}\ and\ \bibinfo {author} {\bibfnamefont {N.}~\bibnamefont {Bonini}},\ }\href {https://doi.org/10.1103/PhysRevB.98.201201} {\bibfield  {journal} {\bibinfo  {journal} {Phys. Rev. B}\ }\textbf {\bibinfo {volume} {98}},\ \bibinfo {pages} {201201(R)} (\bibinfo {year} {2018})}\BibitemShut {NoStop}%
\bibitem [{\citenamefont {Ma}\ \emph {et~al.}(2018{\natexlab{a}})\citenamefont {Ma}, \citenamefont {Nissimagoudar},\ and\ \citenamefont {Li}}]{Ma2018a}%
  \BibitemOpen
  \bibfield  {author} {\bibinfo {author} {\bibfnamefont {J.}~\bibnamefont {Ma}}, \bibinfo {author} {\bibfnamefont {A.~S.}\ \bibnamefont {Nissimagoudar}},\ and\ \bibinfo {author} {\bibfnamefont {W.}~\bibnamefont {Li}},\ }\href {https://doi.org/10.1103/PhysRevB.97.045201} {\bibfield  {journal} {\bibinfo  {journal} {Phys. Rev. B}\ }\textbf {\bibinfo {volume} {97}},\ \bibinfo {pages} {045201} (\bibinfo {year} {2018}{\natexlab{a}})}\BibitemShut {NoStop}%
\bibitem [{\citenamefont {Ma}\ \emph {et~al.}(2018{\natexlab{b}})\citenamefont {Ma}, \citenamefont {Chen},\ and\ \citenamefont {Li}}]{Ma2018b}%
  \BibitemOpen
  \bibfield  {author} {\bibinfo {author} {\bibfnamefont {J.}~\bibnamefont {Ma}}, \bibinfo {author} {\bibfnamefont {Y.}~\bibnamefont {Chen}},\ and\ \bibinfo {author} {\bibfnamefont {W.}~\bibnamefont {Li}},\ }\href {https://doi.org/10.1103/PhysRevB.97.205207} {\bibfield  {journal} {\bibinfo  {journal} {Phys. Rev. B}\ }\textbf {\bibinfo {volume} {97}},\ \bibinfo {pages} {205207} (\bibinfo {year} {2018}{\natexlab{b}})}\BibitemShut {NoStop}%
\bibitem [{\citenamefont {Liu}\ \emph {et~al.}(2017)\citenamefont {Liu}, \citenamefont {Zhou}, \citenamefont {Liao}, \citenamefont {Singh},\ and\ \citenamefont {Chen}}]{Liu2017}%
  \BibitemOpen
  \bibfield  {author} {\bibinfo {author} {\bibfnamefont {T.-H.}\ \bibnamefont {Liu}}, \bibinfo {author} {\bibfnamefont {J.}~\bibnamefont {Zhou}}, \bibinfo {author} {\bibfnamefont {B.}~\bibnamefont {Liao}}, \bibinfo {author} {\bibfnamefont {D.~J.}\ \bibnamefont {Singh}},\ and\ \bibinfo {author} {\bibfnamefont {G.}~\bibnamefont {Chen}},\ }\href {https://doi.org/10.1103/PhysRevB.95.075206} {\bibfield  {journal} {\bibinfo  {journal} {Phys. Rev. B}\ }\textbf {\bibinfo {volume} {95}},\ \bibinfo {pages} {075206} (\bibinfo {year} {2017})}\BibitemShut {NoStop}%
\bibitem [{\citenamefont {Liu}\ \emph {et~al.}(2018)\citenamefont {Liu}, \citenamefont {Song}, \citenamefont {Meroueh}, \citenamefont {Ding}, \citenamefont {Song}, \citenamefont {Zhou}, \citenamefont {Li},\ and\ \citenamefont {Chen}}]{Liu2018}%
  \BibitemOpen
  \bibfield  {author} {\bibinfo {author} {\bibfnamefont {T.-H.}\ \bibnamefont {Liu}}, \bibinfo {author} {\bibfnamefont {B.}~\bibnamefont {Song}}, \bibinfo {author} {\bibfnamefont {L.}~\bibnamefont {Meroueh}}, \bibinfo {author} {\bibfnamefont {Z.}~\bibnamefont {Ding}}, \bibinfo {author} {\bibfnamefont {Q.}~\bibnamefont {Song}}, \bibinfo {author} {\bibfnamefont {J.}~\bibnamefont {Zhou}}, \bibinfo {author} {\bibfnamefont {M.}~\bibnamefont {Li}},\ and\ \bibinfo {author} {\bibfnamefont {G.}~\bibnamefont {Chen}},\ }\bibfield  {title} {\bibinfo {title} {Simultaneously high electron and hole mobilities in cubic boron-v compounds: Bp, bas, and bsb},\ }\href {https://doi.org/10.1103/PhysRevB.98.081203} {\bibfield  {journal} {\bibinfo  {journal} {Phys. Rev. B}\ }\textbf {\bibinfo {volume} {98}},\ \bibinfo {pages} {081203(R)} (\bibinfo {year} {2018})}\BibitemShut {NoStop}%
\bibitem [{\citenamefont {Li}(2015)}]{Li2015}%
  \BibitemOpen
  \bibfield  {author} {\bibinfo {author} {\bibfnamefont {W.}~\bibnamefont {Li}},\ }\href {https://doi.org/10.1103/PhysRevB.92.075405} {\bibfield  {journal} {\bibinfo  {journal} {Phys. Rev. B}\ }\textbf {\bibinfo {volume} {92}},\ \bibinfo {pages} {075405} (\bibinfo {year} {2015})}\BibitemShut {NoStop}%
\bibitem [{\citenamefont {Lee}\ \emph {et~al.}(2018)\citenamefont {Lee}, \citenamefont {Zhou}, \citenamefont {Agapito},\ and\ \citenamefont {Bernardi}}]{Lee2018}%
  \BibitemOpen
  \bibfield  {author} {\bibinfo {author} {\bibfnamefont {N.-E.}\ \bibnamefont {Lee}}, \bibinfo {author} {\bibfnamefont {J.-J.}\ \bibnamefont {Zhou}}, \bibinfo {author} {\bibfnamefont {L.~A.}\ \bibnamefont {Agapito}},\ and\ \bibinfo {author} {\bibfnamefont {M.}~\bibnamefont {Bernardi}},\ }\href {https://doi.org/10.1103/PhysRevB.97.115203} {\bibfield  {journal} {\bibinfo  {journal} {Phys. Rev. B}\ }\textbf {\bibinfo {volume} {97}},\ \bibinfo {pages} {115203} (\bibinfo {year} {2018})}\BibitemShut {NoStop}%
\bibitem [{\citenamefont {Lee}\ \emph {et~al.}(2020)\citenamefont {Lee}, \citenamefont {Zhou}, \citenamefont {Chen},\ and\ \citenamefont {Bernardi}}]{Lee2020}%
  \BibitemOpen
  \bibfield  {author} {\bibinfo {author} {\bibfnamefont {N.~E.}\ \bibnamefont {Lee}}, \bibinfo {author} {\bibfnamefont {J.~J.}\ \bibnamefont {Zhou}}, \bibinfo {author} {\bibfnamefont {H.~Y.}\ \bibnamefont {Chen}},\ and\ \bibinfo {author} {\bibfnamefont {M.}~\bibnamefont {Bernardi}},\ }\href {https://doi.org/10.1038/s41467-020-15339-0} {\bibfield  {journal} {\bibinfo  {journal} {Nat. Commun.}\ }\textbf {\bibinfo {volume} {11}},\ \bibinfo {pages} {1} (\bibinfo {year} {2020})}\BibitemShut {NoStop}%
\bibitem [{\citenamefont {Kang}\ \emph {et~al.}(2019)\citenamefont {Kang}, \citenamefont {Peelaers},\ and\ \citenamefont {Van~de Walle}}]{Kang2019}%
  \BibitemOpen
  \bibfield  {author} {\bibinfo {author} {\bibfnamefont {Y.}~\bibnamefont {Kang}}, \bibinfo {author} {\bibfnamefont {H.}~\bibnamefont {Peelaers}},\ and\ \bibinfo {author} {\bibfnamefont {C.~G.}\ \bibnamefont {Van~de Walle}},\ }\href {https://doi.org/10.1103/PhysRevB.100.121113} {\bibfield  {journal} {\bibinfo  {journal} {Phys. Rev. B}\ }\textbf {\bibinfo {volume} {100}},\ \bibinfo {pages} {121113} (\bibinfo {year} {2019})}\BibitemShut {NoStop}%
\bibitem [{\citenamefont {Jhalani}\ \emph {et~al.}(2020)\citenamefont {Jhalani}, \citenamefont {Zhou}, \citenamefont {Park}, \citenamefont {Dreyer},\ and\ \citenamefont {Bernardi}}]{Jhalani2020}%
  \BibitemOpen
  \bibfield  {author} {\bibinfo {author} {\bibfnamefont {V.~A.}\ \bibnamefont {Jhalani}}, \bibinfo {author} {\bibfnamefont {J.-J.}\ \bibnamefont {Zhou}}, \bibinfo {author} {\bibfnamefont {J.}~\bibnamefont {Park}}, \bibinfo {author} {\bibfnamefont {C.~E.}\ \bibnamefont {Dreyer}},\ and\ \bibinfo {author} {\bibfnamefont {M.}~\bibnamefont {Bernardi}},\ }\href {https://doi.org/10.1103/PhysRevLett.125.136602} {\bibfield  {journal} {\bibinfo  {journal} {Phys. Rev. Lett.}\ }\textbf {\bibinfo {volume} {125}},\ \bibinfo {pages} {136602} (\bibinfo {year} {2020})}\BibitemShut {NoStop}%
\bibitem [{\citenamefont {Cao}\ \emph {et~al.}(2018)\citenamefont {Cao}, \citenamefont {Querales-Flores}, \citenamefont {Murphy}, \citenamefont {Fahy},\ and\ \citenamefont {Savi\ifmmode~\acute{c}\else \'{c}\fi{}}}]{Cao2018}%
  \BibitemOpen
  \bibfield  {author} {\bibinfo {author} {\bibfnamefont {J.}~\bibnamefont {Cao}}, \bibinfo {author} {\bibfnamefont {J.~D.}\ \bibnamefont {Querales-Flores}}, \bibinfo {author} {\bibfnamefont {A.~R.}\ \bibnamefont {Murphy}}, \bibinfo {author} {\bibfnamefont {S.}~\bibnamefont {Fahy}},\ and\ \bibinfo {author} {\bibfnamefont {I.}~\bibnamefont {Savi\ifmmode~\acute{c}\else \'{c}\fi{}}},\ }\href {https://doi.org/10.1103/PhysRevB.98.205202} {\bibfield  {journal} {\bibinfo  {journal} {Phys. Rev. B}\ }\textbf {\bibinfo {volume} {98}},\ \bibinfo {pages} {205202} (\bibinfo {year} {2018})}\BibitemShut {NoStop}%
\bibitem [{\citenamefont {Brunin}\ \emph {et~al.}(2020{\natexlab{a}})\citenamefont {Brunin}, \citenamefont {Miranda}, \citenamefont {Giantomassi}, \citenamefont {Royo}, \citenamefont {Stengel}, \citenamefont {Verstraete}, \citenamefont {Gonze}, \citenamefont {Rignanese},\ and\ \citenamefont {Hautier}}]{Brunin2020b}%
  \BibitemOpen
  \bibfield  {author} {\bibinfo {author} {\bibfnamefont {G.}~\bibnamefont {Brunin}}, \bibinfo {author} {\bibfnamefont {H.~P.~C.}\ \bibnamefont {Miranda}}, \bibinfo {author} {\bibfnamefont {M.}~\bibnamefont {Giantomassi}}, \bibinfo {author} {\bibfnamefont {M.}~\bibnamefont {Royo}}, \bibinfo {author} {\bibfnamefont {M.}~\bibnamefont {Stengel}}, \bibinfo {author} {\bibfnamefont {M.~J.}\ \bibnamefont {Verstraete}}, \bibinfo {author} {\bibfnamefont {X.}~\bibnamefont {Gonze}}, \bibinfo {author} {\bibfnamefont {G.-M.}\ \bibnamefont {Rignanese}},\ and\ \bibinfo {author} {\bibfnamefont {G.}~\bibnamefont {Hautier}},\ }\href {https://doi.org/10.1103/PhysRevB.102.094308} {\bibfield  {journal} {\bibinfo  {journal} {Phys. Rev. B}\ }\textbf {\bibinfo {volume} {102}},\ \bibinfo {pages} {094308} (\bibinfo {year} {2020}{\natexlab{a}})}\BibitemShut {NoStop}%
\bibitem [{\citenamefont {Zhou}\ \emph {et~al.}(2018)\citenamefont {Zhou}, \citenamefont {Hellman},\ and\ \citenamefont {Bernardi}}]{Zhou2018}%
  \BibitemOpen
  \bibfield  {author} {\bibinfo {author} {\bibfnamefont {J.~J.}\ \bibnamefont {Zhou}}, \bibinfo {author} {\bibfnamefont {O.}~\bibnamefont {Hellman}},\ and\ \bibinfo {author} {\bibfnamefont {M.}~\bibnamefont {Bernardi}},\ }\href {https://doi.org/10.1103/PHYSREVLETT.121.226603} {\bibfield  {journal} {\bibinfo  {journal} {Phys. Rev. Lett.}\ }\textbf {\bibinfo {volume} {121}},\ \bibinfo {pages} {226603} (\bibinfo {year} {2018})}\BibitemShut {NoStop}%
\bibitem [{\citenamefont {Zhou}\ and\ \citenamefont {Bernardi}(2019)}]{Zhou2019}%
  \BibitemOpen
  \bibfield  {author} {\bibinfo {author} {\bibfnamefont {J.~J.}\ \bibnamefont {Zhou}}\ and\ \bibinfo {author} {\bibfnamefont {M.}~\bibnamefont {Bernardi}},\ }\href {https://doi.org/10.1103/PHYSREVRESEARCH.1.033138} {\bibfield  {journal} {\bibinfo  {journal} {Phys. Rev. Res.}\ }\textbf {\bibinfo {volume} {1}},\ \bibinfo {pages} {033138} (\bibinfo {year} {2019})}\BibitemShut {NoStop}%
\bibitem [{\citenamefont {Ma}\ \emph {et~al.}(2022)\citenamefont {Ma}, \citenamefont {Li},\ and\ \citenamefont {Luo}}]{Ma2022}%
  \BibitemOpen
  \bibfield  {author} {\bibinfo {author} {\bibfnamefont {J.}~\bibnamefont {Ma}}, \bibinfo {author} {\bibfnamefont {W.}~\bibnamefont {Li}},\ and\ \bibinfo {author} {\bibfnamefont {X.}~\bibnamefont {Luo}},\ }\href {https://doi.org/10.1103/PhysRevB.106.045201} {\bibfield  {journal} {\bibinfo  {journal} {Phys. Rev. B}\ }\textbf {\bibinfo {volume} {106}},\ \bibinfo {pages} {045201} (\bibinfo {year} {2022})}\BibitemShut {NoStop}%
\bibitem [{\citenamefont {Wang}\ \emph {et~al.}(2023)\citenamefont {Wang}, \citenamefont {Wang}, \citenamefont {Chi}, \citenamefont {Ouyang}, \citenamefont {Guo}, \citenamefont {Yang},\ and\ \citenamefont {Chen}}]{Wang2023}%
  \BibitemOpen
  \bibfield  {author} {\bibinfo {author} {\bibfnamefont {Q.}~\bibnamefont {Wang}}, \bibinfo {author} {\bibfnamefont {C.}~\bibnamefont {Wang}}, \bibinfo {author} {\bibfnamefont {C.}~\bibnamefont {Chi}}, \bibinfo {author} {\bibfnamefont {N.}~\bibnamefont {Ouyang}}, \bibinfo {author} {\bibfnamefont {R.}~\bibnamefont {Guo}}, \bibinfo {author} {\bibfnamefont {N.}~\bibnamefont {Yang}},\ and\ \bibinfo {author} {\bibfnamefont {Y.}~\bibnamefont {Chen}},\ }\href {https://doi.org/10.1103/PhysRevB.108.115435} {\bibfield  {journal} {\bibinfo  {journal} {Phys. Rev. B}\ }\textbf {\bibinfo {volume} {108}},\ \bibinfo {pages} {115435} (\bibinfo {year} {2023})}\BibitemShut {NoStop}%
\bibitem [{\citenamefont {Shen}\ \emph {et~al.}(2023)\citenamefont {Shen}, \citenamefont {Dai}, \citenamefont {Xiao}, \citenamefont {Hadaeghi}, \citenamefont {Xie}, \citenamefont {Weidenkaff}, \citenamefont {Tadano},\ and\ \citenamefont {Zhang}}]{Shen2023}%
  \BibitemOpen
  \bibfield  {author} {\bibinfo {author} {\bibfnamefont {C.}~\bibnamefont {Shen}}, \bibinfo {author} {\bibfnamefont {M.}~\bibnamefont {Dai}}, \bibinfo {author} {\bibfnamefont {X.}~\bibnamefont {Xiao}}, \bibinfo {author} {\bibfnamefont {N.}~\bibnamefont {Hadaeghi}}, \bibinfo {author} {\bibfnamefont {W.}~\bibnamefont {Xie}}, \bibinfo {author} {\bibfnamefont {A.}~\bibnamefont {Weidenkaff}}, \bibinfo {author} {\bibfnamefont {T.}~\bibnamefont {Tadano}},\ and\ \bibinfo {author} {\bibfnamefont {H.}~\bibnamefont {Zhang}},\ }\href {https://doi.org/https://doi.org/10.1016/j.mtphys.2023.101059} {\bibfield  {journal} {\bibinfo  {journal} {Mater. Today Phys.}\ }\textbf {\bibinfo {volume} {34}},\ \bibinfo {pages} {101059} (\bibinfo {year} {2023})}\BibitemShut {NoStop}%
\bibitem [{\citenamefont {Wang}\ \emph {et~al.}(2016)\citenamefont {Wang}, \citenamefont {Walker}, \citenamefont {Tamai}, \citenamefont {Wang}, \citenamefont {Ristic}, \citenamefont {Bruno}, \citenamefont {Torre}, \citenamefont {Riccò}, \citenamefont {Plumb}, \citenamefont {Shi}, \citenamefont {Hlawenka}, \citenamefont {Sánchez-Barriga}, \citenamefont {Varykhalov}, \citenamefont {Kim}, \citenamefont {Hoesch}, \citenamefont {King}, \citenamefont {Meevasana}, \citenamefont {Diebold}, \citenamefont {Mesot}, \citenamefont {Moritz}, \citenamefont {Devereaux}, \citenamefont {Radovic},\ and\ \citenamefont {Baumberger}}]{Wang2016}%
  \BibitemOpen
  \bibfield  {author} {\bibinfo {author} {\bibfnamefont {Z.}~\bibnamefont {Wang}}, \bibinfo {author} {\bibfnamefont {S.~M.}\ \bibnamefont {Walker}}, \bibinfo {author} {\bibfnamefont {A.}~\bibnamefont {Tamai}}, \bibinfo {author} {\bibfnamefont {Y.}~\bibnamefont {Wang}}, \bibinfo {author} {\bibfnamefont {Z.}~\bibnamefont {Ristic}}, \bibinfo {author} {\bibfnamefont {F.~Y.}\ \bibnamefont {Bruno}}, \bibinfo {author} {\bibfnamefont {A.~D.~L.}\ \bibnamefont {Torre}}, \bibinfo {author} {\bibfnamefont {S.}~\bibnamefont {Riccò}}, \bibinfo {author} {\bibfnamefont {N.~C.}\ \bibnamefont {Plumb}}, \bibinfo {author} {\bibfnamefont {M.}~\bibnamefont {Shi}}, \bibinfo {author} {\bibfnamefont {P.}~\bibnamefont {Hlawenka}}, \bibinfo {author} {\bibfnamefont {J.}~\bibnamefont {Sánchez-Barriga}}, \bibinfo {author} {\bibfnamefont {A.}~\bibnamefont {Varykhalov}}, \bibinfo {author} {\bibfnamefont {T.~K.}\ \bibnamefont {Kim}}, \bibinfo {author} {\bibfnamefont {M.}~\bibnamefont {Hoesch}}, \bibinfo {author} {\bibfnamefont {P.~D.}\
  \bibnamefont {King}}, \bibinfo {author} {\bibfnamefont {W.}~\bibnamefont {Meevasana}}, \bibinfo {author} {\bibfnamefont {U.}~\bibnamefont {Diebold}}, \bibinfo {author} {\bibfnamefont {J.}~\bibnamefont {Mesot}}, \bibinfo {author} {\bibfnamefont {B.}~\bibnamefont {Moritz}}, \bibinfo {author} {\bibfnamefont {T.~P.}\ \bibnamefont {Devereaux}}, \bibinfo {author} {\bibfnamefont {M.}~\bibnamefont {Radovic}},\ and\ \bibinfo {author} {\bibfnamefont {F.}~\bibnamefont {Baumberger}},\ }\href {https://doi.org/10.1038/nmat4623} {\bibfield  {journal} {\bibinfo  {journal} {Nat. Mater.}\ }\textbf {\bibinfo {volume} {15}},\ \bibinfo {pages} {835} (\bibinfo {year} {2016})}\BibitemShut {NoStop}%
\bibitem [{\citenamefont {van Mechelen}\ \emph {et~al.}(2008)\citenamefont {van Mechelen}, \citenamefont {van~der Marel}, \citenamefont {Grimaldi}, \citenamefont {Kuzmenko}, \citenamefont {Armitage}, \citenamefont {Reyren}, \citenamefont {Hagemann},\ and\ \citenamefont {Mazin}}]{Mechelen2008}%
  \BibitemOpen
  \bibfield  {author} {\bibinfo {author} {\bibfnamefont {J.~L.~M.}\ \bibnamefont {van Mechelen}}, \bibinfo {author} {\bibfnamefont {D.}~\bibnamefont {van~der Marel}}, \bibinfo {author} {\bibfnamefont {C.}~\bibnamefont {Grimaldi}}, \bibinfo {author} {\bibfnamefont {A.~B.}\ \bibnamefont {Kuzmenko}}, \bibinfo {author} {\bibfnamefont {N.~P.}\ \bibnamefont {Armitage}}, \bibinfo {author} {\bibfnamefont {N.}~\bibnamefont {Reyren}}, \bibinfo {author} {\bibfnamefont {H.}~\bibnamefont {Hagemann}},\ and\ \bibinfo {author} {\bibfnamefont {I.~I.}\ \bibnamefont {Mazin}},\ }\href {https://doi.org/10.1103/PhysRevLett.100.226403} {\bibfield  {journal} {\bibinfo  {journal} {Phys. Rev. Lett.}\ }\textbf {\bibinfo {volume} {100}},\ \bibinfo {pages} {226403} (\bibinfo {year} {2008})}\BibitemShut {NoStop}%
\bibitem [{\citenamefont {Klimin}\ \emph {et~al.}(2020)\citenamefont {Klimin}, \citenamefont {Tempere}, \citenamefont {Devreese}, \citenamefont {Franchini},\ and\ \citenamefont {Kresse}}]{Klimin2020}%
  \BibitemOpen
  \bibfield  {author} {\bibinfo {author} {\bibfnamefont {S.}~\bibnamefont {Klimin}}, \bibinfo {author} {\bibfnamefont {J.}~\bibnamefont {Tempere}}, \bibinfo {author} {\bibfnamefont {J.~T.}\ \bibnamefont {Devreese}}, \bibinfo {author} {\bibfnamefont {C.}~\bibnamefont {Franchini}},\ and\ \bibinfo {author} {\bibfnamefont {G.}~\bibnamefont {Kresse}},\ }\bibfield  {title} {\bibinfo {title} {Optical response of an interacting polaron gas in strongly polar crystals},\ }\bibfield  {journal} {\bibinfo  {journal} {Applied Sciences}\ }\textbf {\bibinfo {volume} {10}},\ \href {https://doi.org/10.3390/app10062059} {10.3390/app10062059} (\bibinfo {year} {2020})\BibitemShut {NoStop}%
\bibitem [{\citenamefont {Santander-Syro}\ \emph {et~al.}(2012)\citenamefont {Santander-Syro}, \citenamefont {Bareille}, \citenamefont {Fortuna}, \citenamefont {Copie}, \citenamefont {Gabay}, \citenamefont {Bertran}, \citenamefont {Taleb-Ibrahimi}, \citenamefont {Le~F\`evre}, \citenamefont {Herranz}, \citenamefont {Reyren}, \citenamefont {Bibes}, \citenamefont {Barth\'el\'emy}, \citenamefont {Lecoeur}, \citenamefont {Guevara},\ and\ \citenamefont {Rozenberg}}]{Santander-Syro2012}%
  \BibitemOpen
  \bibfield  {author} {\bibinfo {author} {\bibfnamefont {A.~F.}\ \bibnamefont {Santander-Syro}}, \bibinfo {author} {\bibfnamefont {C.}~\bibnamefont {Bareille}}, \bibinfo {author} {\bibfnamefont {F.}~\bibnamefont {Fortuna}}, \bibinfo {author} {\bibfnamefont {O.}~\bibnamefont {Copie}}, \bibinfo {author} {\bibfnamefont {M.}~\bibnamefont {Gabay}}, \bibinfo {author} {\bibfnamefont {F.}~\bibnamefont {Bertran}}, \bibinfo {author} {\bibfnamefont {A.}~\bibnamefont {Taleb-Ibrahimi}}, \bibinfo {author} {\bibfnamefont {P.}~\bibnamefont {Le~F\`evre}}, \bibinfo {author} {\bibfnamefont {G.}~\bibnamefont {Herranz}}, \bibinfo {author} {\bibfnamefont {N.}~\bibnamefont {Reyren}}, \bibinfo {author} {\bibfnamefont {M.}~\bibnamefont {Bibes}}, \bibinfo {author} {\bibfnamefont {A.}~\bibnamefont {Barth\'el\'emy}}, \bibinfo {author} {\bibfnamefont {P.}~\bibnamefont {Lecoeur}}, \bibinfo {author} {\bibfnamefont {J.}~\bibnamefont {Guevara}},\ and\ \bibinfo {author} {\bibfnamefont {M.~J.}\ \bibnamefont {Rozenberg}},\ }\href
  {https://doi.org/10.1103/PhysRevB.86.121107} {\bibfield  {journal} {\bibinfo  {journal} {Phys. Rev. B}\ }\textbf {\bibinfo {volume} {86}},\ \bibinfo {pages} {121107(R)} (\bibinfo {year} {2012})}\BibitemShut {NoStop}%
\bibitem [{\citenamefont {Cain}\ \emph {et~al.}(2013)\citenamefont {Cain}, \citenamefont {Kajdos},\ and\ \citenamefont {Stemmer}}]{Cain2013}%
  \BibitemOpen
  \bibfield  {author} {\bibinfo {author} {\bibfnamefont {T.~A.}\ \bibnamefont {Cain}}, \bibinfo {author} {\bibfnamefont {A.~P.}\ \bibnamefont {Kajdos}},\ and\ \bibinfo {author} {\bibfnamefont {S.}~\bibnamefont {Stemmer}},\ }\href {https://doi.org/10.1063/1.4804182} {\bibfield  {journal} {\bibinfo  {journal} {Appl. Phys. Lett.}\ }\textbf {\bibinfo {volume} {102}},\ \bibinfo {pages} {182101} (\bibinfo {year} {2013})}\BibitemShut {NoStop}%
\bibitem [{\citenamefont {Wemple}(1965)}]{Wemple1965}%
  \BibitemOpen
  \bibfield  {author} {\bibinfo {author} {\bibfnamefont {S.~H.}\ \bibnamefont {Wemple}},\ }\href {https://doi.org/10.1103/PhysRev.137.A1575} {\bibfield  {journal} {\bibinfo  {journal} {Phys. Rev.}\ }\textbf {\bibinfo {volume} {137}},\ \bibinfo {pages} {A1575} (\bibinfo {year} {1965})}\BibitemShut {NoStop}%
\bibitem [{\citenamefont {Hooton}(1955)}]{Hooton1955}%
  \BibitemOpen
  \bibfield  {author} {\bibinfo {author} {\bibfnamefont {D.}~\bibnamefont {Hooton}},\ }\href {https://doi.org/10.1080/14786440408520575} {\bibfield  {journal} {\bibinfo  {journal} {London Edinburgh Philos. Mag. \& J. Sci.}\ }\textbf {\bibinfo {volume} {46}},\ \bibinfo {pages} {422} (\bibinfo {year} {1955})}\BibitemShut {NoStop}%
\bibitem [{\citenamefont {Errea}\ \emph {et~al.}(2013)\citenamefont {Errea}, \citenamefont {Calandra},\ and\ \citenamefont {Mauri}}]{Errea2013}%
  \BibitemOpen
  \bibfield  {author} {\bibinfo {author} {\bibfnamefont {I.}~\bibnamefont {Errea}}, \bibinfo {author} {\bibfnamefont {M.}~\bibnamefont {Calandra}},\ and\ \bibinfo {author} {\bibfnamefont {F.}~\bibnamefont {Mauri}},\ }\href {https://doi.org/10.1103/PhysRevLett.111.177002} {\bibfield  {journal} {\bibinfo  {journal} {Phys. Rev. Lett.}\ }\textbf {\bibinfo {volume} {111}},\ \bibinfo {pages} {177002} (\bibinfo {year} {2013})}\BibitemShut {NoStop}%
\bibitem [{\citenamefont {Errea}\ \emph {et~al.}(2014)\citenamefont {Errea}, \citenamefont {Calandra},\ and\ \citenamefont {Mauri}}]{Errea2014}%
  \BibitemOpen
  \bibfield  {author} {\bibinfo {author} {\bibfnamefont {I.}~\bibnamefont {Errea}}, \bibinfo {author} {\bibfnamefont {M.}~\bibnamefont {Calandra}},\ and\ \bibinfo {author} {\bibfnamefont {F.}~\bibnamefont {Mauri}},\ }\href {https://doi.org/10.1103/PhysRevB.89.064302} {\bibfield  {journal} {\bibinfo  {journal} {Phys. Rev. B}\ }\textbf {\bibinfo {volume} {89}},\ \bibinfo {pages} {064302} (\bibinfo {year} {2014})}\BibitemShut {NoStop}%
\bibitem [{\citenamefont {Bianco}\ \emph {et~al.}(2017)\citenamefont {Bianco}, \citenamefont {Errea}, \citenamefont {Paulatto}, \citenamefont {Calandra},\ and\ \citenamefont {Mauri}}]{Bianco2017}%
  \BibitemOpen
  \bibfield  {author} {\bibinfo {author} {\bibfnamefont {R.}~\bibnamefont {Bianco}}, \bibinfo {author} {\bibfnamefont {I.}~\bibnamefont {Errea}}, \bibinfo {author} {\bibfnamefont {L.}~\bibnamefont {Paulatto}}, \bibinfo {author} {\bibfnamefont {M.}~\bibnamefont {Calandra}},\ and\ \bibinfo {author} {\bibfnamefont {F.}~\bibnamefont {Mauri}},\ }\href {https://doi.org/10.1103/PhysRevB.96.014111} {\bibfield  {journal} {\bibinfo  {journal} {Phys. Rev. B}\ }\textbf {\bibinfo {volume} {96}},\ \bibinfo {pages} {014111} (\bibinfo {year} {2017})}\BibitemShut {NoStop}%
\bibitem [{\citenamefont {Monacelli}\ \emph {et~al.}(2018)\citenamefont {Monacelli}, \citenamefont {Errea}, \citenamefont {Calandra},\ and\ \citenamefont {Mauri}}]{Monacelli2018}%
  \BibitemOpen
  \bibfield  {author} {\bibinfo {author} {\bibfnamefont {L.}~\bibnamefont {Monacelli}}, \bibinfo {author} {\bibfnamefont {I.}~\bibnamefont {Errea}}, \bibinfo {author} {\bibfnamefont {M.}~\bibnamefont {Calandra}},\ and\ \bibinfo {author} {\bibfnamefont {F.}~\bibnamefont {Mauri}},\ }\href {https://doi.org/10.1103/PhysRevB.98.024106} {\bibfield  {journal} {\bibinfo  {journal} {Phys. Rev. B}\ }\textbf {\bibinfo {volume} {98}},\ \bibinfo {pages} {024106} (\bibinfo {year} {2018})}\BibitemShut {NoStop}%
\bibitem [{\citenamefont {Monacelli}\ \emph {et~al.}(2021)\citenamefont {Monacelli}, \citenamefont {Bianco}, \citenamefont {Cherubini}, \citenamefont {Calandra}, \citenamefont {Errea},\ and\ \citenamefont {Mauri}}]{Monacelli2021}%
  \BibitemOpen
  \bibfield  {author} {\bibinfo {author} {\bibfnamefont {L.}~\bibnamefont {Monacelli}}, \bibinfo {author} {\bibfnamefont {R.}~\bibnamefont {Bianco}}, \bibinfo {author} {\bibfnamefont {M.}~\bibnamefont {Cherubini}}, \bibinfo {author} {\bibfnamefont {M.}~\bibnamefont {Calandra}}, \bibinfo {author} {\bibfnamefont {I.}~\bibnamefont {Errea}},\ and\ \bibinfo {author} {\bibfnamefont {F.}~\bibnamefont {Mauri}},\ }\href {https://doi.org/10.1088/1361-648X/AC066B} {\bibfield  {journal} {\bibinfo  {journal} {J. Condens. Matter Phys}\ }\textbf {\bibinfo {volume} {33}},\ \bibinfo {pages} {363001} (\bibinfo {year} {2021})}\BibitemShut {NoStop}%
\bibitem [{\citenamefont {Verdi}\ \emph {et~al.}(2023)\citenamefont {Verdi}, \citenamefont {Ranalli}, \citenamefont {Franchini},\ and\ \citenamefont {Kresse}}]{Verdi2023}%
  \BibitemOpen
  \bibfield  {author} {\bibinfo {author} {\bibfnamefont {C.}~\bibnamefont {Verdi}}, \bibinfo {author} {\bibfnamefont {L.}~\bibnamefont {Ranalli}}, \bibinfo {author} {\bibfnamefont {C.}~\bibnamefont {Franchini}},\ and\ \bibinfo {author} {\bibfnamefont {G.}~\bibnamefont {Kresse}},\ }\href {https://doi.org/10.1103/PhysRevMaterials.7.L030801} {\bibfield  {journal} {\bibinfo  {journal} {Phys. Rev. Mater.}\ }\textbf {\bibinfo {volume} {7}},\ \bibinfo {pages} {L030801} (\bibinfo {year} {2023})}\BibitemShut {NoStop}%
\bibitem [{\citenamefont {Ranalli}\ \emph {et~al.}(2023)\citenamefont {Ranalli}, \citenamefont {Verdi}, \citenamefont {Monacelli}, \citenamefont {Kresse}, \citenamefont {Calandra},\ and\ \citenamefont {Franchini}}]{Ranalli2023}%
  \BibitemOpen
  \bibfield  {author} {\bibinfo {author} {\bibfnamefont {L.}~\bibnamefont {Ranalli}}, \bibinfo {author} {\bibfnamefont {C.}~\bibnamefont {Verdi}}, \bibinfo {author} {\bibfnamefont {L.}~\bibnamefont {Monacelli}}, \bibinfo {author} {\bibfnamefont {G.}~\bibnamefont {Kresse}}, \bibinfo {author} {\bibfnamefont {M.}~\bibnamefont {Calandra}},\ and\ \bibinfo {author} {\bibfnamefont {C.}~\bibnamefont {Franchini}},\ }\href {https://doi.org/10.1002/QUTE.202200131} {\bibfield  {journal} {\bibinfo  {journal} {Adv. Quantum Technol.}\ ,\ \bibinfo {pages} {2200131}} (\bibinfo {year} {2023})}\BibitemShut {NoStop}%
\bibitem [{\citenamefont {Zacharias}\ \emph {et~al.}(2023{\natexlab{a}})\citenamefont {Zacharias}, \citenamefont {Volonakis}, \citenamefont {Giustino},\ and\ \citenamefont {Even}}]{Zacharias2023}%
  \BibitemOpen
  \bibfield  {author} {\bibinfo {author} {\bibfnamefont {M.}~\bibnamefont {Zacharias}}, \bibinfo {author} {\bibfnamefont {G.}~\bibnamefont {Volonakis}}, \bibinfo {author} {\bibfnamefont {F.}~\bibnamefont {Giustino}},\ and\ \bibinfo {author} {\bibfnamefont {J.}~\bibnamefont {Even}},\ }\href {https://doi.org/10.1038/s41524-023-01089-2} {\bibfield  {journal} {\bibinfo  {journal} {Npj Comput. Mater.}\ }\textbf {\bibinfo {volume} {9}},\ \bibinfo {pages} {1} (\bibinfo {year} {2023}{\natexlab{a}})}\BibitemShut {NoStop}%
\bibitem [{\citenamefont {Jinnouchi}\ \emph {et~al.}(2019)\citenamefont {Jinnouchi}, \citenamefont {Karsai},\ and\ \citenamefont {Kresse}}]{Jinnouchi2019}%
  \BibitemOpen
  \bibfield  {author} {\bibinfo {author} {\bibfnamefont {R.}~\bibnamefont {Jinnouchi}}, \bibinfo {author} {\bibfnamefont {F.}~\bibnamefont {Karsai}},\ and\ \bibinfo {author} {\bibfnamefont {G.}~\bibnamefont {Kresse}},\ }\href {https://doi.org/10.1103/PhysRevB.100.014105} {\bibfield  {journal} {\bibinfo  {journal} {Phys. Rev. B}\ }\textbf {\bibinfo {volume} {100}},\ \bibinfo {pages} {014105} (\bibinfo {year} {2019})}\BibitemShut {NoStop}%
\bibitem [{\citenamefont {Ponc\'e}\ \emph {et~al.}(2016)\citenamefont {Ponc\'e}, \citenamefont {Margine}, \citenamefont {Verdi},\ and\ \citenamefont {Giustino}}]{ponce2016}%
  \BibitemOpen
  \bibfield  {author} {\bibinfo {author} {\bibfnamefont {S.}~\bibnamefont {Ponc\'e}}, \bibinfo {author} {\bibfnamefont {E.~R.}\ \bibnamefont {Margine}}, \bibinfo {author} {\bibfnamefont {C.}~\bibnamefont {Verdi}},\ and\ \bibinfo {author} {\bibfnamefont {F.}~\bibnamefont {Giustino}},\ }\href {https://doi.org/10.1016/J.CPC.2016.07.028} {\bibfield  {journal} {\bibinfo  {journal} {Comput. Phys. Commun.}\ }\textbf {\bibinfo {volume} {209}},\ \bibinfo {pages} {116} (\bibinfo {year} {2016})}\BibitemShut {NoStop}%
\bibitem [{\citenamefont {Giustino}\ \emph {et~al.}(2007)\citenamefont {Giustino}, \citenamefont {Cohen},\ and\ \citenamefont {Louie}}]{Giustino2007}%
  \BibitemOpen
  \bibfield  {author} {\bibinfo {author} {\bibfnamefont {F.}~\bibnamefont {Giustino}}, \bibinfo {author} {\bibfnamefont {M.~L.}\ \bibnamefont {Cohen}},\ and\ \bibinfo {author} {\bibfnamefont {S.~G.}\ \bibnamefont {Louie}},\ }\href {https://doi.org/10.1103/PHYSREVB.76.165108} {\bibfield  {journal} {\bibinfo  {journal} {Phys. Rev. B}\ }\textbf {\bibinfo {volume} {76}},\ \bibinfo {pages} {165108} (\bibinfo {year} {2007})}\BibitemShut {NoStop}%
\bibitem [{\citenamefont {Verdi}\ and\ \citenamefont {Giustino}(2015)}]{Verdi2015}%
  \BibitemOpen
  \bibfield  {author} {\bibinfo {author} {\bibfnamefont {C.}~\bibnamefont {Verdi}}\ and\ \bibinfo {author} {\bibfnamefont {F.}~\bibnamefont {Giustino}},\ }\href {https://doi.org/10.1103/PHYSREVLETT.115.176401} {\bibfield  {journal} {\bibinfo  {journal} {Phys. Rev. Lett.}\ }\textbf {\bibinfo {volume} {115}},\ \bibinfo {pages} {176401} (\bibinfo {year} {2015})}\BibitemShut {NoStop}%
\bibitem [{\citenamefont {Lee}\ \emph {et~al.}(2023)\citenamefont {Lee}, \citenamefont {Ponc\'e}, \citenamefont {Bushick}, \citenamefont {Hajinazar}, \citenamefont {Lafuente-Bartolome}, \citenamefont {Leveillee}, \citenamefont {Lian}, \citenamefont {Lihm}, \citenamefont {Macheda}, \citenamefont {Mori}, \citenamefont {Paudyal}, \citenamefont {Sio}, \citenamefont {Tiwari}, \citenamefont {Zacharias}, \citenamefont {Zhang}, \citenamefont {Bonini}, \citenamefont {Kioupakis}, \citenamefont {Margine},\ and\ \citenamefont {Giustino}}]{Lee2023}%
  \BibitemOpen
  \bibfield  {author} {\bibinfo {author} {\bibfnamefont {H.}~\bibnamefont {Lee}}, \bibinfo {author} {\bibfnamefont {S.}~\bibnamefont {Ponc\'e}}, \bibinfo {author} {\bibfnamefont {K.}~\bibnamefont {Bushick}}, \bibinfo {author} {\bibfnamefont {S.}~\bibnamefont {Hajinazar}}, \bibinfo {author} {\bibfnamefont {J.}~\bibnamefont {Lafuente-Bartolome}}, \bibinfo {author} {\bibfnamefont {J.}~\bibnamefont {Leveillee}}, \bibinfo {author} {\bibfnamefont {C.}~\bibnamefont {Lian}}, \bibinfo {author} {\bibfnamefont {J.~M.}\ \bibnamefont {Lihm}}, \bibinfo {author} {\bibfnamefont {F.}~\bibnamefont {Macheda}}, \bibinfo {author} {\bibfnamefont {H.}~\bibnamefont {Mori}}, \bibinfo {author} {\bibfnamefont {H.}~\bibnamefont {Paudyal}}, \bibinfo {author} {\bibfnamefont {W.~H.}\ \bibnamefont {Sio}}, \bibinfo {author} {\bibfnamefont {S.}~\bibnamefont {Tiwari}}, \bibinfo {author} {\bibfnamefont {M.}~\bibnamefont {Zacharias}}, \bibinfo {author} {\bibfnamefont {X.}~\bibnamefont {Zhang}}, \bibinfo {author} {\bibfnamefont {N.}~\bibnamefont
  {Bonini}}, \bibinfo {author} {\bibfnamefont {E.}~\bibnamefont {Kioupakis}}, \bibinfo {author} {\bibfnamefont {E.~R.}\ \bibnamefont {Margine}},\ and\ \bibinfo {author} {\bibfnamefont {F.}~\bibnamefont {Giustino}},\ }\href {https://doi.org/10.1038/S41524-023-01107-3} {\bibfield  {journal} {\bibinfo  {journal} {Npj Comput. Mater.}\ }\textbf {\bibinfo {volume} {9}},\ \bibinfo {pages} {1} (\bibinfo {year} {2023})}\BibitemShut {NoStop}%
\bibitem [{SM()}]{SM}%
  \BibitemOpen
  \href@noop {} {}\bibinfo {note} {See Supplemental Material at [URL], which includes Refs.~\cite{Zhou2018,Verdi2023,Monacelli2021}, for a summary of the theory of phonon-limited drift mobility, details on the anharmonic corrections to the electron-phonon matrix elements, wannierization of the disordered bands, MLFF construction and SSCHA calculations, additional mobility results, and Figs. S1–S10.}\BibitemShut {Stop}%
\bibitem [{\citenamefont {Giannozzi}\ \emph {et~al.}(2009)\citenamefont {Giannozzi}, \citenamefont {Baroni}, \citenamefont {Bonini}, \citenamefont {Calandra}, \citenamefont {Car}, \citenamefont {Cavazzoni}, \citenamefont {Ceresoli}, \citenamefont {Chiarotti}, \citenamefont {Cococcioni}, \citenamefont {Dabo}, \citenamefont {Corso}, \citenamefont {de~Gironcoli}, \citenamefont {Fabris}, \citenamefont {Fratesi}, \citenamefont {Gebauer}, \citenamefont {Gerstmann}, \citenamefont {Gougoussis}, \citenamefont {Kokalj}, \citenamefont {Lazzeri}, \citenamefont {Martin-Samos}, \citenamefont {Marzari}, \citenamefont {Mauri}, \citenamefont {Mazzarello}, \citenamefont {Paolini}, \citenamefont {Pasquarello}, \citenamefont {Paulatto}, \citenamefont {Sbraccia}, \citenamefont {Scandolo}, \citenamefont {Sclauzero}, \citenamefont {Seitsonen}, \citenamefont {Smogunov}, \citenamefont {Umari},\ and\ \citenamefont {Wentzcovitch}}]{Giannozzi2009}%
  \BibitemOpen
  \bibfield  {author} {\bibinfo {author} {\bibfnamefont {P.}~\bibnamefont {Giannozzi}}, \bibinfo {author} {\bibfnamefont {S.}~\bibnamefont {Baroni}}, \bibinfo {author} {\bibfnamefont {N.}~\bibnamefont {Bonini}}, \bibinfo {author} {\bibfnamefont {M.}~\bibnamefont {Calandra}}, \bibinfo {author} {\bibfnamefont {R.}~\bibnamefont {Car}}, \bibinfo {author} {\bibfnamefont {C.}~\bibnamefont {Cavazzoni}}, \bibinfo {author} {\bibfnamefont {D.}~\bibnamefont {Ceresoli}}, \bibinfo {author} {\bibfnamefont {G.~L.}\ \bibnamefont {Chiarotti}}, \bibinfo {author} {\bibfnamefont {M.}~\bibnamefont {Cococcioni}}, \bibinfo {author} {\bibfnamefont {I.}~\bibnamefont {Dabo}}, \bibinfo {author} {\bibfnamefont {A.~D.}\ \bibnamefont {Corso}}, \bibinfo {author} {\bibfnamefont {S.}~\bibnamefont {de~Gironcoli}}, \bibinfo {author} {\bibfnamefont {S.}~\bibnamefont {Fabris}}, \bibinfo {author} {\bibfnamefont {G.}~\bibnamefont {Fratesi}}, \bibinfo {author} {\bibfnamefont {R.}~\bibnamefont {Gebauer}}, \bibinfo {author} {\bibfnamefont
  {U.}~\bibnamefont {Gerstmann}}, \bibinfo {author} {\bibfnamefont {C.}~\bibnamefont {Gougoussis}}, \bibinfo {author} {\bibfnamefont {A.}~\bibnamefont {Kokalj}}, \bibinfo {author} {\bibfnamefont {M.}~\bibnamefont {Lazzeri}}, \bibinfo {author} {\bibfnamefont {L.}~\bibnamefont {Martin-Samos}}, \bibinfo {author} {\bibfnamefont {N.}~\bibnamefont {Marzari}}, \bibinfo {author} {\bibfnamefont {F.}~\bibnamefont {Mauri}}, \bibinfo {author} {\bibfnamefont {R.}~\bibnamefont {Mazzarello}}, \bibinfo {author} {\bibfnamefont {S.}~\bibnamefont {Paolini}}, \bibinfo {author} {\bibfnamefont {A.}~\bibnamefont {Pasquarello}}, \bibinfo {author} {\bibfnamefont {L.}~\bibnamefont {Paulatto}}, \bibinfo {author} {\bibfnamefont {C.}~\bibnamefont {Sbraccia}}, \bibinfo {author} {\bibfnamefont {S.}~\bibnamefont {Scandolo}}, \bibinfo {author} {\bibfnamefont {G.}~\bibnamefont {Sclauzero}}, \bibinfo {author} {\bibfnamefont {A.~P.}\ \bibnamefont {Seitsonen}}, \bibinfo {author} {\bibfnamefont {A.}~\bibnamefont {Smogunov}}, \bibinfo {author}
  {\bibfnamefont {P.}~\bibnamefont {Umari}},\ and\ \bibinfo {author} {\bibfnamefont {R.~M.}\ \bibnamefont {Wentzcovitch}},\ }\href {https://doi.org/10.1088/0953-8984/21/39/395502} {\bibfield  {journal} {\bibinfo  {journal} {J. Phys.: Condens. Matter}\ }\textbf {\bibinfo {volume} {21}},\ \bibinfo {pages} {395502} (\bibinfo {year} {2009})}\BibitemShut {NoStop}%
\bibitem [{\citenamefont {Giannozzi}\ \emph {et~al.}(2017)\citenamefont {Giannozzi}, \citenamefont {Andreussi}, \citenamefont {Brumme}, \citenamefont {Bunau}, \citenamefont {Nardelli}, \citenamefont {Calandra}, \citenamefont {Car}, \citenamefont {Cavazzoni}, \citenamefont {Ceresoli}, \citenamefont {Cococcioni}, \citenamefont {Colonna}, \citenamefont {Carnimeo}, \citenamefont {Corso}, \citenamefont {de~Gironcoli}, \citenamefont {Delugas}, \citenamefont {DiStasio}, \citenamefont {Ferretti}, \citenamefont {Floris}, \citenamefont {Fratesi}, \citenamefont {Fugallo}, \citenamefont {Gebauer}, \citenamefont {Gerstmann}, \citenamefont {Giustino}, \citenamefont {Gorni}, \citenamefont {Jia}, \citenamefont {Kawamura}, \citenamefont {Ko}, \citenamefont {Kokalj}, \citenamefont {Küçükbenli}, \citenamefont {Lazzeri}, \citenamefont {Marsili}, \citenamefont {Marzari}, \citenamefont {Mauri}, \citenamefont {Nguyen}, \citenamefont {Nguyen}, \citenamefont {de-la Roza}, \citenamefont {Paulatto}, \citenamefont {Ponc\'e},
  \citenamefont {Rocca}, \citenamefont {Sabatini}, \citenamefont {Santra}, \citenamefont {Schlipf}, \citenamefont {Seitsonen}, \citenamefont {Smogunov}, \citenamefont {Timrov}, \citenamefont {Thonhauser}, \citenamefont {Umari}, \citenamefont {Vast}, \citenamefont {Wu},\ and\ \citenamefont {Baroni}}]{Giannozzi2017}%
  \BibitemOpen
  \bibfield  {author} {\bibinfo {author} {\bibfnamefont {P.}~\bibnamefont {Giannozzi}}, \bibinfo {author} {\bibfnamefont {O.}~\bibnamefont {Andreussi}}, \bibinfo {author} {\bibfnamefont {T.}~\bibnamefont {Brumme}}, \bibinfo {author} {\bibfnamefont {O.}~\bibnamefont {Bunau}}, \bibinfo {author} {\bibfnamefont {M.~B.}\ \bibnamefont {Nardelli}}, \bibinfo {author} {\bibfnamefont {M.}~\bibnamefont {Calandra}}, \bibinfo {author} {\bibfnamefont {R.}~\bibnamefont {Car}}, \bibinfo {author} {\bibfnamefont {C.}~\bibnamefont {Cavazzoni}}, \bibinfo {author} {\bibfnamefont {D.}~\bibnamefont {Ceresoli}}, \bibinfo {author} {\bibfnamefont {M.}~\bibnamefont {Cococcioni}}, \bibinfo {author} {\bibfnamefont {N.}~\bibnamefont {Colonna}}, \bibinfo {author} {\bibfnamefont {I.}~\bibnamefont {Carnimeo}}, \bibinfo {author} {\bibfnamefont {A.~D.}\ \bibnamefont {Corso}}, \bibinfo {author} {\bibfnamefont {S.}~\bibnamefont {de~Gironcoli}}, \bibinfo {author} {\bibfnamefont {P.}~\bibnamefont {Delugas}}, \bibinfo {author} {\bibfnamefont
  {R.~A.}\ \bibnamefont {DiStasio}}, \bibinfo {author} {\bibfnamefont {A.}~\bibnamefont {Ferretti}}, \bibinfo {author} {\bibfnamefont {A.}~\bibnamefont {Floris}}, \bibinfo {author} {\bibfnamefont {G.}~\bibnamefont {Fratesi}}, \bibinfo {author} {\bibfnamefont {G.}~\bibnamefont {Fugallo}}, \bibinfo {author} {\bibfnamefont {R.}~\bibnamefont {Gebauer}}, \bibinfo {author} {\bibfnamefont {U.}~\bibnamefont {Gerstmann}}, \bibinfo {author} {\bibfnamefont {F.}~\bibnamefont {Giustino}}, \bibinfo {author} {\bibfnamefont {T.}~\bibnamefont {Gorni}}, \bibinfo {author} {\bibfnamefont {J.}~\bibnamefont {Jia}}, \bibinfo {author} {\bibfnamefont {M.}~\bibnamefont {Kawamura}}, \bibinfo {author} {\bibfnamefont {H.-Y.}\ \bibnamefont {Ko}}, \bibinfo {author} {\bibfnamefont {A.}~\bibnamefont {Kokalj}}, \bibinfo {author} {\bibfnamefont {E.}~\bibnamefont {Küçükbenli}}, \bibinfo {author} {\bibfnamefont {M.}~\bibnamefont {Lazzeri}}, \bibinfo {author} {\bibfnamefont {M.}~\bibnamefont {Marsili}}, \bibinfo {author} {\bibfnamefont
  {N.}~\bibnamefont {Marzari}}, \bibinfo {author} {\bibfnamefont {F.}~\bibnamefont {Mauri}}, \bibinfo {author} {\bibfnamefont {N.~L.}\ \bibnamefont {Nguyen}}, \bibinfo {author} {\bibfnamefont {H.-V.}\ \bibnamefont {Nguyen}}, \bibinfo {author} {\bibfnamefont {A.~O.}\ \bibnamefont {de-la Roza}}, \bibinfo {author} {\bibfnamefont {L.}~\bibnamefont {Paulatto}}, \bibinfo {author} {\bibfnamefont {S.}~\bibnamefont {Ponc\'e}}, \bibinfo {author} {\bibfnamefont {D.}~\bibnamefont {Rocca}}, \bibinfo {author} {\bibfnamefont {R.}~\bibnamefont {Sabatini}}, \bibinfo {author} {\bibfnamefont {B.}~\bibnamefont {Santra}}, \bibinfo {author} {\bibfnamefont {M.}~\bibnamefont {Schlipf}}, \bibinfo {author} {\bibfnamefont {A.~P.}\ \bibnamefont {Seitsonen}}, \bibinfo {author} {\bibfnamefont {A.}~\bibnamefont {Smogunov}}, \bibinfo {author} {\bibfnamefont {I.}~\bibnamefont {Timrov}}, \bibinfo {author} {\bibfnamefont {T.}~\bibnamefont {Thonhauser}}, \bibinfo {author} {\bibfnamefont {P.}~\bibnamefont {Umari}}, \bibinfo {author}
  {\bibfnamefont {N.}~\bibnamefont {Vast}}, \bibinfo {author} {\bibfnamefont {X.}~\bibnamefont {Wu}},\ and\ \bibinfo {author} {\bibfnamefont {S.}~\bibnamefont {Baroni}},\ }\href {https://doi.org/10.1088/1361-648X/AA8F79} {\bibfield  {journal} {\bibinfo  {journal} {J. Phys.: Condens. Matter}\ }\textbf {\bibinfo {volume} {29}},\ \bibinfo {pages} {465901} (\bibinfo {year} {2017})}\BibitemShut {NoStop}%
\bibitem [{\citenamefont {Giustino}(2017)}]{Giustino2017}%
  \BibitemOpen
  \bibfield  {author} {\bibinfo {author} {\bibfnamefont {F.}~\bibnamefont {Giustino}},\ }\href {https://doi.org/10.1103/RevModPhys.89.015003} {\bibfield  {journal} {\bibinfo  {journal} {Rev. Mod. Phys.}\ }\textbf {\bibinfo {volume} {89}},\ \bibinfo {pages} {015003} (\bibinfo {year} {2017})}\BibitemShut {NoStop}%
\bibitem [{\citenamefont {Pizzi}\ \emph {et~al.}(2020)\citenamefont {Pizzi}, \citenamefont {Vitale}, \citenamefont {Arita}, \citenamefont {Blügel}, \citenamefont {Freimuth}, \citenamefont {G{\'{e}}ranton}, \citenamefont {Gibertini}, \citenamefont {Gresch}, \citenamefont {Johnson}, \citenamefont {Koretsune}, \citenamefont {Iba{\~{n}}ez-Azpiroz}, \citenamefont {Lee}, \citenamefont {Lihm}, \citenamefont {Marchand}, \citenamefont {Marrazzo}, \citenamefont {Mokrousov}, \citenamefont {Mustafa}, \citenamefont {Nohara}, \citenamefont {Nomura}, \citenamefont {Paulatto}, \citenamefont {Ponc\'e}, \citenamefont {Ponweiser}, \citenamefont {Qiao}, \citenamefont {Thöle}, \citenamefont {Tsirkin}, \citenamefont {Wierzbowska}, \citenamefont {Marzari}, \citenamefont {Vanderbilt}, \citenamefont {Souza}, \citenamefont {Mostofi},\ and\ \citenamefont {Yates}}]{Pizzi2020}%
  \BibitemOpen
  \bibfield  {author} {\bibinfo {author} {\bibfnamefont {G.}~\bibnamefont {Pizzi}}, \bibinfo {author} {\bibfnamefont {V.}~\bibnamefont {Vitale}}, \bibinfo {author} {\bibfnamefont {R.}~\bibnamefont {Arita}}, \bibinfo {author} {\bibfnamefont {S.}~\bibnamefont {Blügel}}, \bibinfo {author} {\bibfnamefont {F.}~\bibnamefont {Freimuth}}, \bibinfo {author} {\bibfnamefont {G.}~\bibnamefont {G{\'{e}}ranton}}, \bibinfo {author} {\bibfnamefont {M.}~\bibnamefont {Gibertini}}, \bibinfo {author} {\bibfnamefont {D.}~\bibnamefont {Gresch}}, \bibinfo {author} {\bibfnamefont {C.}~\bibnamefont {Johnson}}, \bibinfo {author} {\bibfnamefont {T.}~\bibnamefont {Koretsune}}, \bibinfo {author} {\bibfnamefont {J.}~\bibnamefont {Iba{\~{n}}ez-Azpiroz}}, \bibinfo {author} {\bibfnamefont {H.}~\bibnamefont {Lee}}, \bibinfo {author} {\bibfnamefont {J.-M.}\ \bibnamefont {Lihm}}, \bibinfo {author} {\bibfnamefont {D.}~\bibnamefont {Marchand}}, \bibinfo {author} {\bibfnamefont {A.}~\bibnamefont {Marrazzo}}, \bibinfo {author} {\bibfnamefont
  {Y.}~\bibnamefont {Mokrousov}}, \bibinfo {author} {\bibfnamefont {J.~I.}\ \bibnamefont {Mustafa}}, \bibinfo {author} {\bibfnamefont {Y.}~\bibnamefont {Nohara}}, \bibinfo {author} {\bibfnamefont {Y.}~\bibnamefont {Nomura}}, \bibinfo {author} {\bibfnamefont {L.}~\bibnamefont {Paulatto}}, \bibinfo {author} {\bibfnamefont {S.}~\bibnamefont {Ponc\'e}}, \bibinfo {author} {\bibfnamefont {T.}~\bibnamefont {Ponweiser}}, \bibinfo {author} {\bibfnamefont {J.}~\bibnamefont {Qiao}}, \bibinfo {author} {\bibfnamefont {F.}~\bibnamefont {Thöle}}, \bibinfo {author} {\bibfnamefont {S.~S.}\ \bibnamefont {Tsirkin}}, \bibinfo {author} {\bibfnamefont {M.}~\bibnamefont {Wierzbowska}}, \bibinfo {author} {\bibfnamefont {N.}~\bibnamefont {Marzari}}, \bibinfo {author} {\bibfnamefont {D.}~\bibnamefont {Vanderbilt}}, \bibinfo {author} {\bibfnamefont {I.}~\bibnamefont {Souza}}, \bibinfo {author} {\bibfnamefont {A.~A.}\ \bibnamefont {Mostofi}},\ and\ \bibinfo {author} {\bibfnamefont {J.~R.}\ \bibnamefont {Yates}},\ }\href
  {https://doi.org/10.1088/1361-648x/ab51ff} {\bibfield  {journal} {\bibinfo  {journal} {J. Condens. Matter Phys.}\ }\textbf {\bibinfo {volume} {32}},\ \bibinfo {pages} {165902} (\bibinfo {year} {2020})}\BibitemShut {NoStop}%
\bibitem [{\citenamefont {Wannier}(1937)}]{wannier1937}%
  \BibitemOpen
  \bibfield  {author} {\bibinfo {author} {\bibfnamefont {G.~H.}\ \bibnamefont {Wannier}},\ }\href {https://doi.org/10.1103/PhysRev.52.191} {\bibfield  {journal} {\bibinfo  {journal} {Phys. Rev.}\ }\textbf {\bibinfo {volume} {52}},\ \bibinfo {pages} {191} (\bibinfo {year} {1937})}\BibitemShut {NoStop}%
\bibitem [{\citenamefont {Marzari}\ and\ \citenamefont {Vanderbilt}(1997)}]{Marzari1997}%
  \BibitemOpen
  \bibfield  {author} {\bibinfo {author} {\bibfnamefont {N.}~\bibnamefont {Marzari}}\ and\ \bibinfo {author} {\bibfnamefont {D.}~\bibnamefont {Vanderbilt}},\ }\href {https://doi.org/10.1103/PhysRevB.56.12847} {\bibfield  {journal} {\bibinfo  {journal} {Phys. Rev. B}\ }\textbf {\bibinfo {volume} {56}},\ \bibinfo {pages} {12847} (\bibinfo {year} {1997})}\BibitemShut {NoStop}%
\bibitem [{\citenamefont {Perdew}\ \emph {et~al.}(1996)\citenamefont {Perdew}, \citenamefont {Burke},\ and\ \citenamefont {Ernzerhof}}]{Perdew1997}%
  \BibitemOpen
  \bibfield  {author} {\bibinfo {author} {\bibfnamefont {J.~P.}\ \bibnamefont {Perdew}}, \bibinfo {author} {\bibfnamefont {K.}~\bibnamefont {Burke}},\ and\ \bibinfo {author} {\bibfnamefont {M.}~\bibnamefont {Ernzerhof}},\ }\href {https://doi.org/10.1103/PhysRevLett.77.3865} {\bibfield  {journal} {\bibinfo  {journal} {Phys. Rev. Lett.}\ }\textbf {\bibinfo {volume} {77}},\ \bibinfo {pages} {3865} (\bibinfo {year} {1996})}\BibitemShut {NoStop}%
\bibitem [{\citenamefont {Hamann}(2013)}]{Hamann2013}%
  \BibitemOpen
  \bibfield  {author} {\bibinfo {author} {\bibfnamefont {D.~R.}\ \bibnamefont {Hamann}},\ }\href {https://doi.org/10.1103/PhysRevB.88.085117} {\bibfield  {journal} {\bibinfo  {journal} {Phys. Rev. B}\ }\textbf {\bibinfo {volume} {88}},\ \bibinfo {pages} {085117} (\bibinfo {year} {2013})}\BibitemShut {NoStop}%
\bibitem [{\citenamefont {Kresse}\ and\ \citenamefont {Hafner}(1993)}]{Kresse1993}%
  \BibitemOpen
  \bibfield  {author} {\bibinfo {author} {\bibfnamefont {G.}~\bibnamefont {Kresse}}\ and\ \bibinfo {author} {\bibfnamefont {J.}~\bibnamefont {Hafner}},\ }\href {https://doi.org/10.1103/PhysRevB.47.558} {\bibfield  {journal} {\bibinfo  {journal} {Phys. Rev. B}\ }\textbf {\bibinfo {volume} {47}},\ \bibinfo {pages} {558} (\bibinfo {year} {1993})}\BibitemShut {NoStop}%
\bibitem [{\citenamefont {Kresse}\ and\ \citenamefont {Furthm{\"{u}}ller}(1996)}]{Kresse1996}%
  \BibitemOpen
  \bibfield  {author} {\bibinfo {author} {\bibfnamefont {G.}~\bibnamefont {Kresse}}\ and\ \bibinfo {author} {\bibfnamefont {J.}~\bibnamefont {Furthm{\"{u}}ller}},\ }\href {https://doi.org/10.1016/0927-0256(96)00008-0} {\bibfield  {journal} {\bibinfo  {journal} {Comput. Mater. Sci.}\ }\textbf {\bibinfo {volume} {6}},\ \bibinfo {pages} {15} (\bibinfo {year} {1996})}\BibitemShut {NoStop}%
\bibitem [{\citenamefont {Hoover}\ \emph {et~al.}(1982)\citenamefont {Hoover}, \citenamefont {Ladd},\ and\ \citenamefont {Moran}}]{Hoover1982}%
  \BibitemOpen
  \bibfield  {author} {\bibinfo {author} {\bibfnamefont {W.~G.}\ \bibnamefont {Hoover}}, \bibinfo {author} {\bibfnamefont {A.~J.~C.}\ \bibnamefont {Ladd}},\ and\ \bibinfo {author} {\bibfnamefont {B.}~\bibnamefont {Moran}},\ }\href {https://doi.org/10.1103/PhysRevLett.48.1818} {\bibfield  {journal} {\bibinfo  {journal} {Phys. Rev. Lett.}\ }\textbf {\bibinfo {volume} {48}},\ \bibinfo {pages} {1818} (\bibinfo {year} {1982})}\BibitemShut {NoStop}%
\bibitem [{\citenamefont {Sun}\ \emph {et~al.}(2015)\citenamefont {Sun}, \citenamefont {Ruzsinszky},\ and\ \citenamefont {Perdew}}]{Sun2015}%
  \BibitemOpen
  \bibfield  {author} {\bibinfo {author} {\bibfnamefont {J.}~\bibnamefont {Sun}}, \bibinfo {author} {\bibfnamefont {A.}~\bibnamefont {Ruzsinszky}},\ and\ \bibinfo {author} {\bibfnamefont {J.~P.}\ \bibnamefont {Perdew}},\ }\href {https://doi.org/10.1103/PhysRevLett.115.036402} {\bibfield  {journal} {\bibinfo  {journal} {Phys. Rev. Lett.}\ }\textbf {\bibinfo {volume} {115}},\ \bibinfo {pages} {036402} (\bibinfo {year} {2015})}\BibitemShut {NoStop}%
\bibitem [{\citenamefont {Kresse}\ and\ \citenamefont {Joubert}(1999)}]{Kresse1999}%
  \BibitemOpen
  \bibfield  {author} {\bibinfo {author} {\bibfnamefont {G.}~\bibnamefont {Kresse}}\ and\ \bibinfo {author} {\bibfnamefont {D.}~\bibnamefont {Joubert}},\ }\href {https://doi.org/10.1103/PhysRevB.59.1758} {\bibfield  {journal} {\bibinfo  {journal} {Phys. Rev. B}\ }\textbf {\bibinfo {volume} {59}},\ \bibinfo {pages} {1758} (\bibinfo {year} {1999})}\BibitemShut {NoStop}%
\bibitem [{\citenamefont {Pick}\ \emph {et~al.}(1970)\citenamefont {Pick}, \citenamefont {Cohen},\ and\ \citenamefont {Martin}}]{Pick1970}%
  \BibitemOpen
  \bibfield  {author} {\bibinfo {author} {\bibfnamefont {R.~M.}\ \bibnamefont {Pick}}, \bibinfo {author} {\bibfnamefont {M.~H.}\ \bibnamefont {Cohen}},\ and\ \bibinfo {author} {\bibfnamefont {R.~M.}\ \bibnamefont {Martin}},\ }\href {https://doi.org/10.1103/PhysRevB.1.910} {\bibfield  {journal} {\bibinfo  {journal} {Phys. Rev. B}\ }\textbf {\bibinfo {volume} {1}},\ \bibinfo {pages} {910} (\bibinfo {year} {1970})}\BibitemShut {NoStop}%
\bibitem [{\citenamefont {Zhao}\ \emph {et~al.}(2020{\natexlab{b}})\citenamefont {Zhao}, \citenamefont {Dalpian}, \citenamefont {Wang},\ and\ \citenamefont {Zunger}}]{Zhao2020a}%
  \BibitemOpen
  \bibfield  {author} {\bibinfo {author} {\bibfnamefont {X.-G.}\ \bibnamefont {Zhao}}, \bibinfo {author} {\bibfnamefont {G.~M.}\ \bibnamefont {Dalpian}}, \bibinfo {author} {\bibfnamefont {Z.}~\bibnamefont {Wang}},\ and\ \bibinfo {author} {\bibfnamefont {A.}~\bibnamefont {Zunger}},\ }\bibfield  {title} {\bibinfo {title} {Polymorphous nature of cubic halide perovskites},\ }\href {https://doi.org/10.1103/PhysRevB.101.155137} {\bibfield  {journal} {\bibinfo  {journal} {Phys. Rev. B}\ }\textbf {\bibinfo {volume} {101}},\ \bibinfo {pages} {155137} (\bibinfo {year} {2020}{\natexlab{b}})}\BibitemShut {NoStop}%
\bibitem [{\citenamefont {Zacharias}\ \emph {et~al.}(2023{\natexlab{b}})\citenamefont {Zacharias}, \citenamefont {Volonakis}, \citenamefont {Giustino},\ and\ \citenamefont {Even}}]{Zacharias2023prb}%
  \BibitemOpen
  \bibfield  {author} {\bibinfo {author} {\bibfnamefont {M.}~\bibnamefont {Zacharias}}, \bibinfo {author} {\bibfnamefont {G.}~\bibnamefont {Volonakis}}, \bibinfo {author} {\bibfnamefont {F.}~\bibnamefont {Giustino}},\ and\ \bibinfo {author} {\bibfnamefont {J.}~\bibnamefont {Even}},\ }\href {https://doi.org/10.1103/PhysRevB.108.035155} {\bibfield  {journal} {\bibinfo  {journal} {Phys. Rev. B}\ }\textbf {\bibinfo {volume} {108}},\ \bibinfo {pages} {035155} (\bibinfo {year} {2023}{\natexlab{b}})}\BibitemShut {NoStop}%
\bibitem [{\citenamefont {Popescu}\ and\ \citenamefont {Zunger}(2012)}]{Popescu2012}%
  \BibitemOpen
  \bibfield  {author} {\bibinfo {author} {\bibfnamefont {V.}~\bibnamefont {Popescu}}\ and\ \bibinfo {author} {\bibfnamefont {A.}~\bibnamefont {Zunger}},\ }\href {https://doi.org/10.1103/PhysRevB.85.085201} {\bibfield  {journal} {\bibinfo  {journal} {Phys. Rev. B}\ }\textbf {\bibinfo {volume} {85}},\ \bibinfo {pages} {085201} (\bibinfo {year} {2012})}\BibitemShut {NoStop}%
\bibitem [{\citenamefont {Zacharias}\ and\ \citenamefont {Giustino}(2020)}]{Zacharias2020}%
  \BibitemOpen
  \bibfield  {author} {\bibinfo {author} {\bibfnamefont {M.}~\bibnamefont {Zacharias}}\ and\ \bibinfo {author} {\bibfnamefont {F.}~\bibnamefont {Giustino}},\ }\href {https://doi.org/10.1103/PhysRevResearch.2.013357} {\bibfield  {journal} {\bibinfo  {journal} {Phys. Rev. Res.}\ }\textbf {\bibinfo {volume} {2}},\ \bibinfo {pages} {013357} (\bibinfo {year} {2020})}\BibitemShut {NoStop}%
\bibitem [{\citenamefont {Servoin}\ \emph {et~al.}(1980)\citenamefont {Servoin}, \citenamefont {Luspin},\ and\ \citenamefont {Gervais}}]{Servoin1980}%
  \BibitemOpen
  \bibfield  {author} {\bibinfo {author} {\bibfnamefont {J.~L.}\ \bibnamefont {Servoin}}, \bibinfo {author} {\bibfnamefont {Y.}~\bibnamefont {Luspin}},\ and\ \bibinfo {author} {\bibfnamefont {F.}~\bibnamefont {Gervais}},\ }\href {https://doi.org/10.1103/PhysRevB.22.5501} {\bibfield  {journal} {\bibinfo  {journal} {Phys. Rev. B}\ }\textbf {\bibinfo {volume} {22}},\ \bibinfo {pages} {5501} (\bibinfo {year} {1980})}\BibitemShut {NoStop}%
\bibitem [{\citenamefont {Stirling}(1972)}]{Stirling1972}%
  \BibitemOpen
  \bibfield  {author} {\bibinfo {author} {\bibfnamefont {W.~G.}\ \bibnamefont {Stirling}},\ }\href {https://doi.org/10.1088/0022-3719/5/19/005} {\bibfield  {journal} {\bibinfo  {journal} {J. Phys. C: Solid State Phys.}\ }\textbf {\bibinfo {volume} {5}},\ \bibinfo {pages} {2711} (\bibinfo {year} {1972})}\BibitemShut {NoStop}%
\bibitem [{\citenamefont {Ponc\'e}\ \emph {et~al.}(2019{\natexlab{c}})\citenamefont {Ponc\'e}, \citenamefont {Schlipf},\ and\ \citenamefont {Giustino}}]{ponce2019c}%
  \BibitemOpen
  \bibfield  {author} {\bibinfo {author} {\bibfnamefont {S.}~\bibnamefont {Ponc\'e}}, \bibinfo {author} {\bibfnamefont {M.}~\bibnamefont {Schlipf}},\ and\ \bibinfo {author} {\bibfnamefont {F.}~\bibnamefont {Giustino}},\ }\href {https://doi.org/10.1021/acsenergylett.8b02346} {\bibfield  {journal} {\bibinfo  {journal} {ACS Energy Lett.}\ }\textbf {\bibinfo {volume} {4}},\ \bibinfo {pages} {456} (\bibinfo {year} {2019}{\natexlab{c}})}\BibitemShut {NoStop}%
\bibitem [{\citenamefont {Himmetoglu}\ and\ \citenamefont {Janotti}(2016)}]{Himmetoglu2016}%
  \BibitemOpen
  \bibfield  {author} {\bibinfo {author} {\bibfnamefont {B.}~\bibnamefont {Himmetoglu}}\ and\ \bibinfo {author} {\bibfnamefont {A.}~\bibnamefont {Janotti}},\ }\href {https://doi.org/10.1088/0953-8984/28/6/065502} {\bibfield  {journal} {\bibinfo  {journal} {J. Condens. Matter Phys.}\ }\textbf {\bibinfo {volume} {28}},\ \bibinfo {pages} {065502} (\bibinfo {year} {2016})}\BibitemShut {NoStop}%
\bibitem [{\citenamefont {Janotti}\ \emph {et~al.}(2011)\citenamefont {Janotti}, \citenamefont {Steiauf},\ and\ \citenamefont {Van~de Walle}}]{janotti2011}%
  \BibitemOpen
  \bibfield  {author} {\bibinfo {author} {\bibfnamefont {A.}~\bibnamefont {Janotti}}, \bibinfo {author} {\bibfnamefont {D.}~\bibnamefont {Steiauf}},\ and\ \bibinfo {author} {\bibfnamefont {C.~G.}\ \bibnamefont {Van~de Walle}},\ }\href {https://doi.org/10.1103/PhysRevB.84.201304} {\bibfield  {journal} {\bibinfo  {journal} {Phys. Rev. B}\ }\textbf {\bibinfo {volume} {84}},\ \bibinfo {pages} {201304(R)} (\bibinfo {year} {2011})}\BibitemShut {NoStop}%
\bibitem [{\citenamefont {Merker}\ \emph {et~al.}(1955)\citenamefont {Merker}, \citenamefont {Plock}, \citenamefont {Levin},\ and\ \citenamefont {Field}}]{Merker1955}%
  \BibitemOpen
  \bibfield  {author} {\bibinfo {author} {\bibfnamefont {L.}~\bibnamefont {Merker}}, \bibinfo {author} {\bibfnamefont {F.~M.}\ \bibnamefont {Plock}}, \bibinfo {author} {\bibfnamefont {S.~B.}\ \bibnamefont {Levin}},\ and\ \bibinfo {author} {\bibfnamefont {N.~J.}\ \bibnamefont {Field}},\ }\href {https://doi.org/10.1364/JOSA.45.000737} {\bibfield  {journal} {\bibinfo  {journal} {J. Opt. Soc. Am.}\ }\textbf {\bibinfo {volume} {45}},\ \bibinfo {pages} {737} (\bibinfo {year} {1955})}\BibitemShut {NoStop}%
\bibitem [{\citenamefont {Perry}\ \emph {et~al.}(1967)\citenamefont {Perry}, \citenamefont {Fertel},\ and\ \citenamefont {McNelly}}]{Perry1967}%
  \BibitemOpen
  \bibfield  {author} {\bibinfo {author} {\bibfnamefont {C.~H.}\ \bibnamefont {Perry}}, \bibinfo {author} {\bibfnamefont {J.~H.}\ \bibnamefont {Fertel}},\ and\ \bibinfo {author} {\bibfnamefont {T.~F.}\ \bibnamefont {McNelly}},\ }\href {https://doi.org/10.1063/1.1712142} {\bibfield  {journal} {\bibinfo  {journal} {J. Chem. Phys}\ }\textbf {\bibinfo {volume} {47}},\ \bibinfo {pages} {1619} (\bibinfo {year} {1967})}\BibitemShut {NoStop}%
\bibitem [{\citenamefont {Fujii}\ and\ \citenamefont {Sakudo}(1976)}]{Fujii1976}%
  \BibitemOpen
  \bibfield  {author} {\bibinfo {author} {\bibfnamefont {Y.}~\bibnamefont {Fujii}}\ and\ \bibinfo {author} {\bibfnamefont {T.}~\bibnamefont {Sakudo}},\ }\href {https://doi.org/10.1143/JPSJ.41.888} {\bibfield  {journal} {\bibinfo  {journal} {J. Phys. Soc. Jpn.}\ }\textbf {\bibinfo {volume} {41}},\ \bibinfo {pages} {888} (\bibinfo {year} {1976})}\BibitemShut {NoStop}%
\bibitem [{\citenamefont {Caruso}\ and\ \citenamefont {Giustino}(2016)}]{giustino2016}%
  \BibitemOpen
  \bibfield  {author} {\bibinfo {author} {\bibfnamefont {F.}~\bibnamefont {Caruso}}\ and\ \bibinfo {author} {\bibfnamefont {F.}~\bibnamefont {Giustino}},\ }\href {https://doi.org/10.1103/PhysRevB.94.115208} {\bibfield  {journal} {\bibinfo  {journal} {Phys. Rev. B}\ }\textbf {\bibinfo {volume} {94}},\ \bibinfo {pages} {115208} (\bibinfo {year} {2016})}\BibitemShut {NoStop}%
\bibitem [{\citenamefont {Brunin}\ \emph {et~al.}(2020{\natexlab{b}})\citenamefont {Brunin}, \citenamefont {Miranda}, \citenamefont {Giantomassi}, \citenamefont {Royo}, \citenamefont {Stengel}, \citenamefont {Verstraete}, \citenamefont {Gonze}, \citenamefont {Rignanese},\ and\ \citenamefont {Hautier}}]{brunin2020c}%
  \BibitemOpen
  \bibfield  {author} {\bibinfo {author} {\bibfnamefont {G.}~\bibnamefont {Brunin}}, \bibinfo {author} {\bibfnamefont {H.~P.~C.}\ \bibnamefont {Miranda}}, \bibinfo {author} {\bibfnamefont {M.}~\bibnamefont {Giantomassi}}, \bibinfo {author} {\bibfnamefont {M.}~\bibnamefont {Royo}}, \bibinfo {author} {\bibfnamefont {M.}~\bibnamefont {Stengel}}, \bibinfo {author} {\bibfnamefont {M.~J.}\ \bibnamefont {Verstraete}}, \bibinfo {author} {\bibfnamefont {X.}~\bibnamefont {Gonze}}, \bibinfo {author} {\bibfnamefont {G.-M.}\ \bibnamefont {Rignanese}},\ and\ \bibinfo {author} {\bibfnamefont {G.}~\bibnamefont {Hautier}},\ }\bibfield  {title} {\bibinfo {title} {Electron-phonon beyond fr\"ohlich: Dynamical quadrupoles in polar and covalent solids},\ }\href {https://doi.org/10.1103/PhysRevLett.125.136601} {\bibfield  {journal} {\bibinfo  {journal} {Phys. Rev. Lett.}\ }\textbf {\bibinfo {volume} {125}},\ \bibinfo {pages} {136601} (\bibinfo {year} {2020}{\natexlab{b}})}\BibitemShut {NoStop}%
\end{thebibliography}
\end{document}